\documentclass[usenatbib]{mn2e}
\usepackage{natbib}
\usepackage{apjfonts}
\usepackage{amsmath}
\usepackage[table]{xcolor}

\usepackage{epsfig}
\usepackage{graphicx}
\usepackage{amssymb}
\usepackage{aas_macros}
\usepackage{color}

\newcommand{\kms}{\,{\rm km \, s^{-1}}}
\newcommand{\kpc}{\,{\rm kpc}}
\newcommand{\mpc}{\,{\rm Mpc}}
\newcommand{\oversim}[2]{\protect{\mbox{\lower0.5ex\vbox{%
   \baselineskip=0pt\lineskip=0.2ex
   \ialign{$\mathsurround=0pt #1\hfil##\hfil$\crcr#2\crcr\sim\crcr}}}}} 

\catcode`\"=\active\let"=\"  

\def\3{{\ss} }

\def\c12{{1\over 2}}

\def\d{{\rm d}}   
   
\def\plusplus{\raise 0.3ex\hbox{${\scriptstyle ++}$}{}}

\def\and{{{\rm M}31}}
\def\mw{{\rm MW}}
\def\lmc{{\rm LMC}}

\def\g{{\rm G}}
\def\a{{\rm A}}
\def\gyr{{\rm Gyr}}
\bibliography{biblio}

\usepackage{array}
\newcolumntype{L}[1]{>{\raggedright\let\newline\\\arraybackslash\hspace{0pt}}m{#1}}
\newcolumntype{C}[1]{>{\centering\let\newline\\\arraybackslash\hspace{0pt}}m{#1}}
\newcolumntype{R}[1]{>{\raggedleft\let\newline\\\arraybackslash\hspace{0pt}}m{#1}}

\begin{document}   
\title[What galaxy masses perturb the local cosmic expansion?]{What galaxy masses perturb the local cosmic expansion?}
\author[J. Pe\~{n}arrubia \& A. Fattahi]{Jorge Pe\~{n}arrubia$^{1}$\thanks{jorpega@roe.ac.uk} \& Azadeh Fattahi$^2$\\
$^1$Institute for Astronomy, University of Edinburgh, Royal Observatory, Blackford Hill, Edinburgh EH9 3HJ, UK\\
$^2$Department of Physics and Astronomy, University of Victoria, PO Box 3055 STN CSC, Victoria, BC, V8W 3P6, Canada\\
}
\maketitle  

\begin{abstract} 
We use 12 cosmological $N$-body simulations of Local Group systems (the {\sc Apostle} models) to inspect the relation between the virial mass of the main haloes ($M_{\rm vir,1}$ and $M_{\rm vir,2}$), the mass derived from the relative motion of the halo pair ($M_{\rm tim}$), and that inferred from the local Hubble flow ($M_{\rm lhf}$). We show that within the Spherical Collapse Model (SCM), the correspondence between the three mass estimates is exact, i.e. $M_{\rm lhf}=M_{\rm tim}=M_{\rm vir,1}+M_{\rm vir,2}$.
However, comparison with {\sc Apostle} simulations reveals that, contrary to what the SCM states, a relatively large fraction of the mass that perturbs the local Hubble flow and drives the relative trajectory of the main galaxies is not contained within $R_{\rm vir}$, and that the amount of ``extra-virial'' mass tends to increase in galaxies with a slow accretion rate. 
In contrast, modelling the peculiar velocities around the Local Group returns an unbiased constraint on the virial mass ratio of the main galaxy pair.
Adopting the outer halo profile found in $N$-body simulations, which scales as $\rho\sim R^{-4}$ at $R\gtrsim R_{\rm vir}$, indicates that the galaxy masses perturbing the local Hubble flow roughly correspond to the asymptotically-convergent (total) masses of the individual haloes. We show that estimates of $M_{\rm vir}$ based on the dynamics of tracers at $R\gg R_{\rm vir}$ require a priori information on the internal matter distribution and the growth rate of the main galaxies, both of which are typically difficult to quantify.
\end{abstract}   

\begin{keywords}
Galaxy: kinematics and dynamics; galaxies: evolution. 
\end{keywords}

\section{Introduction} \label{sec:intro}
In an expanding, flat Universe the rather intuitive concept of {\it galaxy mass} is ill-defined and difficult to infer observationally. Given that the mass of a galaxy is thought to be a key parameter for the early collapse and subsequent virialization of cosmological substructures (e.g. Rees \& Ostriker 1977; White \& Rees 1978; Blumenthal et al. 1984) as well as for galaxy formation and evolutionary processes (see e.g. Kravtsov et al. 2004; Conroy \& Wechsler 2009; Moster et al. 2010; Behroozi et al. 2010; Gao et al. 2011; Laporte et al. 2013; Sawala et al. 2016; Rodr{\'{\i}}guez-Puebla et al.2016), it becomes crucial to quantify the correspondence between the different methods that have been proposed in the literature to infer galaxy masses from the dynamics of visible tracers.

The main difficulty in measuring the mass of a galaxy resides in the limited kinematic information available to an observer in the outskirts of these systems. Indeed, in our current cosmological paradigm galaxies form in the inner-most regions of virialized dark matter structures called {\it haloes} (e.g. Mo, van den Bosch \& White 2010). Within those regions different sorts of kinematic tracers, such as gas, stars and stellar clusters, can be used to constrain the inner mass distribution of a galaxy. However, the amount of baryonic tracers declines progressively at large distances from the halo centre, introducing severe uncertainties in our understanding of the mass distribution in the outer-most regions of galactic haloes, to the point that it becomes extremely challenging to identify the physical {\it edge} of galaxies with the surrounding Universe. 

The very notion of galaxy edge is thwarted by numerical simulations of structure formation, which show that dark matter haloes exhibit smooth density profiles without well-defined boundaries.
These simulations reveal that over-densities in a close-to-homogeneous background induce perturbations in the expansion of the Universe (the so-called Hubble flow) which in principle extend to infinite distances from the source. The gravitational pull slows down the Hubble flow 
around over-dense regions, such that at the {\it turn-around distance} ($R_{\rm ta}$) the (local) expansion of Universe halts. Within this volume we find another scale of interest, that where the number of particles moving away from an over-density becomes roughly equal to that moving towards it. By definition, this region is in {\it dynamical equilibrium} and the spherical radius that contains it is called the {\it virial radius}, $R_{\rm vir}$, which provides a natural boundary between a galaxy halo and its environment. 

A useful, albeit rough, estimate of the size of $R_{\rm vir}$ corresponds to the radius where the free-fall time ($t_{\rm ff}$) is equal to half of the age of the Universe ($t_0$). Within a sphere of mass $M$ and mean density $\langle \rho\rangle =3 M/(4\pi R_{\rm vir}^3)$ the free fall time is $t_{\rm ff}=[3\pi/(32 G\langle \rho\rangle)]^{1/2}=[\pi^2R_{\rm vir}^3/(8 G M)]^{1/2}$. The condition $2t_{\rm ff}=t_0$ thus implies that $R_{\rm vir}=(2GM t_0^2/\pi^2)^{1/3}$. Note that the size of the virialized region expands as $R_{\rm vir}\sim t_0^{2/3}$ at a fixed mass $M$, thus leading to a halo growth that is not driven by accretion of new material or a change in the halo potential, but solely due to a boundary definition based on a timing argument. This effect is usually known as the {\it pseudo-evolution of dark matter haloes} (e.g. Diemand et al. 2007; Cuesta et al. 2008; Diemer et al. 2013; Zemp 2014). Note also that the monotonic expansion of $R_{\rm vir}$ with time suggests that the natural evolution of the cosmic web is towards a set of increasingly isolated haloes in dynamical equilibrium (Busha et al. 2005).

 

Although in principle the virial radius of a galaxy is an observable quantity (i.e. the spherical radius encompassing the volume where the in- and outwards motion of the galaxy constituents balance out), in practice a direct measurement is rarely possible owing to the scarcity of kinematic tracers in the outskirts of dark matter haloes. As a remedy, a widely-used approach to estimate the virial radii of galaxies and galaxy clusters consists of measuring the mass enclosed within the volume populated by kinematic tracers, $M(< r_\star$), which is the {\it extrapolated} to a virial radius, $R_{\rm vir}\gtrsim r_\star$. The extrapolation is typically made upon the adoption of a theoretically-motivated density profile, the most common one being the Navarro, Frenk \& White (1996; 1997) profile, henceforth NFW. However, several studies show that the NFW profile does not describe well the overall shape of dark matter haloes found in cosmological $N$-body simulations (e.g. Prada et al. 2006; Betancort-Rijo et al. 2006; Cuesta et al. 2008).  Deviations from the NFW profile at large radii appear to largely correlate with the rate at which haloes accrete mass (e.g. Ludlow et al. 2013; Diemer \& Kravtsov 2014) and may have an impact on the masses of galaxy clusters derived from X-ray observations (e.g. Avestruz et al. 2014) as well as weak-lensing mass measurements (e.g. Oguri \& Hamana 2011; Becker \& Kravtsov 2011). In addition, Taylor et al. (2015) warns about the dangers of adopting halo profiles found in dark matter-only cosmological models to describe galaxies that have been acted on by baryonic feedback. Using hydrodynamical simulations of Milky Way-like galaxies these authors find that extrapolating $M(<50\kpc)$ measurements out to the virial radius systematically overestimates the halo mass and underestimates the halo concentration.

For local galaxies like the Milky Way and Andromeda extrapolations of the NFW profile yield virial masses that are uncertain by a factor $\sim 2$--$3$. 
Tests against mock data sets reveal that the uncertainty is largely dominated by (i) the unknown impact of baryons on the halo density profile and (ii) the unknown orbital distribution of the tracer particles (e.g. Wang et al. 2015).
For example, Smith et al. (2007) use high-velocity stars from the RAVE survey (Steinmetz et al. 2006; Zwitter et al. 2008) to measure the local escape speed of our Galaxy. The range of values found by these authors ($498<v_{\rm esc}/\kms<608$) leads to a NFW halo with virial mass $M_{\rm vir}=0.85_{-0.29}^{+0.55}\times 10^{12}M_\odot$. 
Xue et al. (2008) extend the mass constraints to $D\lesssim 60\kpc$ by modelling the kinematics of blue horizontal branch stars (BHBs), which yields a similar value $0.8_{-0.2}^{+0.2}\times 10^{12}M_\odot$. Sakamoto et al. (2003) and Battaglia et al. (2005) find $2.5_{-1.0}^{+0.5}\times 10^{12}M_\odot$ and $0.8_{-0.2}^{+1.2}\times 10^{12}M_\odot$, respectively, from a mixed sample of globular cluster, giant stars and satellite galaxies. 
Watkins et al. (2010) applies the Jeans equations to 26 satellite galaxies of the Milky Way and finds a virial mass $0.7$-$3.4\times 10^{12}M_\odot$ depending on the (unknown) orbital distribution of the satellite population. Barber et al. (2014) deal with this uncertainty by directly matching the distribution of subhaloes found in dark matter-only $N$-body simulations against the observed position and velocity of Milky Way dwarf spheroidals. This comparison yields a virial mass in the range $0.6<M_{\rm vir}/(10^{12} M_\odot)<3.1$. 

A number of recent studies avoid profile extrapolation by modelling the dynamics of tracers that lie {\it beyond} the nominal virial radius of the halo. 
For example, by demanding that the total momentum of the Local Group should balance to zero Diaz et al. (2014) estimate the individual masses of the Milky Way and Andromeda to be $M_{\rm MW} = (0.8 \pm 0.5) \times 10^{12} M_\odot$ and $M_{\rm M31} = (1.7 \pm 0.3) \times 10^{12} M_\odot$, respectively. 
Also,  
accreted material reaching its first apocentre after halo collapse
gives rise to a density caustic (Fillmore \& Goldreich 1984; Bertschinger 1985) which can be used to define a halo boundary that is in principle observable (Rines et al. 2013, Patej \& Loeb 2015; More et al. 2016). The outermost caustic manifests itself as a sharp density drop in the halo outskirts at a location known as `back-splash' radius ($R_{\rm sp}$). Unfortunately, the relation between the caustic location and the virial size is very sensitive to the recent mass evolution of the halo, which adds significant uncertainty to the relation between the mass enclosed within the back-splash radius, $M(<R_{\rm sp})$ and $M_{\rm vir}$. Cosmological collisionless simulations show that the back-splash radius varies between $R_{\rm sp}/R_{\rm vir}\sim 0.8$--$1.6$ in haloes with fast or slow accretion rates, respectively (see More et al. 2015 for details).

This paper follows up on the work of Pe\~narrubia et al. (2014, 2016; hereinafter P14 and P16) who construct idealized models of structure growth to describe the (local) cosmic expansion around an isolated, overdense region of the Universe. These models are incorporated into a Bayesian framework to fit orbits to measured distances and velocities of galaxies between 0.8 and 3 Mpc from the Local Group (hereinafter LG). The method returns masses for the Milky Way and M31 which do not rely on {\it a priori} premises on the internal matter distribution in those galaxies nor on equilibrium assumptions (see also Banik \& Zhao 2016 for a similar study).
However, the relation between the mass perturbing the local Hubble flow and the individual virial masses of the main haloes remains unclear to date. In this contribution we use self-consistent cosmological simulations of twelve LG-like systems  (the so-called {\sc Apostle}\footnote{A Project Of Simulating The Local Environement} project, Sawala et al. 2016; Fattahi et al. 2016) to calibrate the correspondence between the mass derived from modelling the local Hubble flow and the individual virial masses of the Milky Way \& M31 analogues.

The arrangement of this paper is as follows. In Section~\ref{sec:mass} we introduce the Spherical Collapse Model, which provides a simple, analytical representation of the non-linear growth of structures in a close-to-homogeneous Universe. Section~\ref{sec:num} outlines the construction of mock data sets using cosmological $N$-body models of the Local Group as well as the method we use to fit peculiar velocities around these systems. Section~\ref{sec:res} inspects the relation between the virial masses of the main Local Group haloes and those inferred from the local Hubble flow. We discuss the validity of the Spherical Collapse Model assumptions in Section~\ref{sec:discussion} and attempt to estimate the virial mass of the Milky Way from the dynamics of nearby ($\lesssim 3\mpc$) galaxies. Our findings are summarized in Section~\ref{sec:sum}.

\section{Galaxy masses in cosmology} \label{sec:mass}
In this work we shall measure and compare three sorts of cosmological masses: (i) the mass enclosed within a spherical overdensity with respect to an evolving density threshold ($M_{\rm vir}$), (ii) the mass derived from the timing argument ($M_{\rm tim}$), and (iii) the mass perturbing the local Hubble flow ($M_{\rm lhf}$). Below it is shown that in idealized models of structure formation the three estimates are equivalent to one another.

\subsection{Virial mass}\label{sec:vir}
Historically, the notion of halo virial mass is connected to the Spherical Collapse Model (henceforth SCM; Gunn \& Gott 1972; Peebles 1980; Peacock 1999), which describes how over-dense regions evolve in an expanding Universe. 
In this model the initial density perturbation follows a top-hat distribution\footnote{Similar results follow for an isolated shell of mass $M$ and initial size $R_{\rm ini}$. Here we focus on the top-hat distribution for convenience.}

$$\rho_{\rm ini}=
\begin{cases}
\bar{\rho}_{\rm ini}(1+\Delta_{\rm ini})& r\le R_{\rm ini}\\
0 & r> R_{\rm ini},
\end{cases}$$
where $\Delta_{\rm ini}\equiv \Delta(t_{\rm ini})$ is a small over-density ($0<\Delta_{\rm ini} \ll 1$) in the mean cosmic density $\bar{\rho}$ measured at a time $t=t_{\rm ini}$ close to the Big Bang ($t=0$). Hence, the mass enclosed within the homogeneous over-density is $M=4\pi\bar{\rho}_{\rm ini}(1+\Delta_{\rm ini})R_{\rm ini}^3/3$. This model provides a useful representation of the over-densities observed in the Cosmic Microwave Background, which have amplitudes $\Delta_{\rm ini}\lesssim 10^{-5}$ at redshift $z\sim 1000$.  At these early times the velocities of the particles within the sphere are assumed to trace the Hubble flow, such that $\d R/\d t\approx R_{\rm ini}H_{\rm ini}$, where $H=(8\pi G \bar\rho/3)^{1/2}$ is the Hubble expansion parameter.

The time evolution of the size of the sphere, $R(t)$, obeys the following equation
\begin{equation}\label{eq:kep}
{\ddot R}=-\frac{G M}{R^2} + H^2\Omega_{\Lambda}R,
\end{equation}
where $\Omega_{\Lambda}=\Lambda c^2/(3H_0^2)$ is the fractional vacuum energy density at $z=0$. 

In an Einstein-de Sitter Universe, ($\Omega_m,\Omega_\Lambda)=(1,0)$, Equation~(\ref{eq:kep}) reduces to a Keplerian equation of motion whose solutions can be parametrized as

\begin{eqnarray}\label{eq:rtkep}
R&=&\frac{R_{\rm ta}}{2}(1-\cos\eta) \\ \nonumber
t&=&\frac{t_{\rm ta}}{\pi}(\eta -\sin \eta),
\end{eqnarray}
where the sub-index `ta' denotes quantities measured at {\it turn-around}, that is the time at which the sphere reaches its maximum size and begins to contract owing to its own self-gravity. At the turn-around time $\eta=\pi$ and 
\begin{equation}\label{eq:rta}
R_{\rm ta}=\bigg[\frac{8GM t_{\rm ta}^2}{\pi^2}\bigg]^{1/3}. 
\end{equation}
After turn-around ($\pi\le \eta\le 2\pi$) the size of the sphere shrinks monotonically with time. Given that the radial orbital period of particles moving in a constant-density medium is independent of $R$, the whole sphere collapses to a point-mass at the time $t_{\rm coll}=2t_{\rm ta}$ (or $\eta=2\pi$). 
This unrealistic prediction arises from some oversimplifying assumptions of the model, namely that all particles move on radial orbits, and that the forces induced by large-scale structures in the Universe do not affect the motion of the particles within $R(t)$. In practice, even perfectly isolated \& spherical self-gravitating systems develop radial orbital instabilities (e.g. McMillan et al. 2006; Pontzen et al. 2015) which transfer angular momentum to particles in a process akin to orbital diffusion in the integral-of-motion space (Pe\~narrubia 2015). 

Numerical experiments show that 
collision-less relaxation drives self-gravitating systems towards a state of {\it virial equilibrium} where the sphere is expected to reach a finite, constant size $R_{\rm vir}$. The SCM assumes that the virialization happens instantaneously at $t=t_{\rm coll}$ and derives the radius $R_{\rm vir}$ from the virial theorem by equating the gravitational energy of a uniform sphere at turn-around, $E=W=-(3/5) GM^2/R_{\rm ta}$, to the virial energy at $t=t_{\rm coll}$, $E=W/2=-(3/10) GM^2/R_{\rm vir}$, such that 
\begin{equation}\label{eq:rvir}
R_{\rm vir}=\frac{1}{2}R_{\rm ta}.
\end{equation}
It is straightforward to show that Equations~(\ref{eq:rta}) and~(\ref{eq:rvir}) recover the virial radius that we obtained in the Introduction via equating the free-fall time of a spherical over-density to the age of the Universe.

It is important to emphasize that the factor $1/2$ on the right-hand side of Equation~(\ref{eq:rvir}) arises from the assumption that top-hat over-densities retain a uniform density throughout their initial expansion and subsequent collapse\footnote{The same result applies to a an isolated shell of mass $M$ and radius $R_{\rm ta}$, the only difference being the proportionality factor in the potential energy, namely $W=-GM^2/R_{\rm ta}$.}. This, however, is at odds with the results of collision-less simulations of structure formation, which show that the inner regions of the potential relax on shorter time-scales that the outer envelope, and that virialized haloes follow a close-to-universal density profile that diverge towards the centre as $\rho(r)\propto r^{-1}$ and falls off at large radii as $\rho(r)\propto r^{-3}$ (Dubinski \& Carlberg 1991; Navarro, Frenk \& White 1996,1997). 
Yet, tests using collisionless $\Lambda$CDM simulations
find that on average $R_{\rm vir}/R_{\rm ta}\approx 0.56$ (Suto et al. 2016), which is remarkably close to the SCM prediction given in Equation~(\ref{eq:rvir}).

In an Einstein-de Sitter cosmology, where the mean density of the Universe evolves as $\bar{\rho}=(6 \pi G t^2)^{-1}$, the density of the sphere at the time of collapse $t_{\rm coll}=2t_{\rm ta}$ can be estimated from Equation~(\ref{eq:rta}) as
\begin{equation}\label{eq:dvir}
1+\Delta=\frac{\rho}{\bar{\rho}}=\frac{9\pi^2}{2}\bigg(\frac{t_{\rm coll}}{t_{\rm ta}}\bigg)^2=18\pi^2\sim 200.
\end{equation}
Thus, the SCM predicts that all virialized haloes have the same mean density within the {\it virial radius} regardless of their mass and the amplitude of the initial density contrast $\Delta_{\rm ini}$. As a result, the value $\Delta=200$ is commonly used to set up the threshold of halo identification in cosmological $N$-body simulations, and by extension to describe galaxy masses in a cosmological context. 

Cosmological $N$-body simulations in a $\Lambda$CDM framework show the value of $\Delta$ depends on the matter density at the time of collapse, $\Omega_m(t_{\rm coll})\equiv \bar{\rho}/\bar{\rho_c}=\bar{\rho}[8\pi G /3H^2(t_{\rm coll})]^{-1}$ (e.g. Lacey \& Cole 1993; Eke et al. 1996). For clarity, in what follows we reserve the suffix `vir' to denote virial quantities with respect to the contrast density value expected in $\Lambda$CDM. 
Setting $\Omega_{m,0}\equiv \Omega_m(z=0)=0.279$ (e.g. Planck collaboration 2014), and using $\Delta_c\equiv (1+\Delta_{\rm vir})\Omega_m=18\pi^2+82x -39x^2$, where $x=\Omega -1$, $\Omega(z)=\Omega_{m,0}(1+z)^3/[\Omega_{m,0}(1+z)^3+\Omega_\Lambda]$ (Bryan \& Norman 1998), yields $\Delta_{\rm vir}\approx 358$ at $z=0$.

 Thus, we find that under the SCM assumptions the mass enclosed within $R_{\rm vir}$ is equal to the mass appearing in Equation~(\ref{eq:kep}), i.e.
\begin{equation}\label{eq:Mvir}
M_{\rm vir}\equiv \frac{4\pi}{3} \bar{\rho}(1+\Delta_{\rm vir})R_{\rm vir}^3=M.
\end{equation}

\subsection{Perturbed Hubble flow}\label{sec:flow}
It was originally pointed out by Lynden-Bell (1981) and Sandage (1986) that Equation~(\ref{eq:kep}) also governs the trajectories of individual, mass-less tracer particles at distances $R\gg R_{\rm vir}$ from of a top-hat over-density (see also Chernin et al. 2009). 

In an Einstein-de Sitter Universe manipulation of Equation~(\ref{eq:rtkep}) yields a simple expression for the {\it perturbed Hubble flow} around a mass $M$ at the time $t=t_0$, which can be written as $V=[n-t_0(GM/R)^{1/2}]R/(m t_0)$, where $n=2^{-3/2}\pi$ and $m\simeq 0.87$ (Lynden-Bell 1981). P14 extends this result to a $\Lambda$CDM cosmology by fitting $m$ and $n$ as a function of $\Omega_\Lambda$, which yields
\begin{equation}\label{eq:vd}
V(R)\approx \big(1.2 + 0.16 \Omega_\Lambda\big)\frac{R}{t_0} - 1.1\bigg(\frac{GM}{R}\bigg)^{1/2}.
\end{equation}
The turn-around radius can be derived from Equation~(\ref{eq:vd}) by setting $v=0$, which yields $R_{\rm ta}=[1.2 GM t^2_0/(1.2+0.16 \Omega_\Lambda)^2]^{1/3}$. Note that the location of the turn-around radius is barely sensitive to the dark energy density, $R_{\rm ta}/R_{\rm ta}(\Omega_\Lambda=0)\simeq (1-0.086\Omega_\Lambda)$. Also, comparison with Equation~(\ref{eq:rta}) shows that the zero-velocity radius derived from Equation~(\ref{eq:vd}) with $\Omega_\Lambda=0$ recovers the turn-around radius predicted by the SCM at a $\sim 1\%$ level.

Equation~(\ref{eq:vd}) can be used to measure $M$ by fitting the observed distance-velocity relation of galaxies at $R\gg R_{\rm vir}$. The correspondence between the mass perturbing the local Hubble flow in Equation~(\ref{eq:vd}), henceforth denoted as $M_{\rm lhf}$, and the virial mass~(\ref{eq:Mvir}) is exact under the SCM assumptions, i.e. 
\begin{equation}\label{eq:mlhf}
M_{\rm lhf}\equiv M=M_{\rm vir}.
\end{equation}

\subsection{Timing argument}\label{sec:timing}
Kahn \& Woltjer (1959) (see also Li \& White 2008; van der Marel 
\& Guhathakurta 2008; van der Marel 2012a,b; Partridge et al. 2013; Gonz\'alez et al. 2014; McLeod et al. 2016; P16) used the fact that Equation~(\ref{eq:kep}) provides an exact description of the relative motion between two isolated point-mass particles in an expanding Universe, ${\bf R}\equiv {\bf R}_{1}-{\bf R}_{2}$, to constrain the combined mass $M=M_{1}+M_{2}$. 
If the age of the Universe is known the mass $M$ can be directly measured from their current separation and relative velocity of the two particles, which is known as the `timing argument'. 

Neglecting the dark energy term in~(\ref{eq:kep}) yields a Keplerian solution to the equations of motion
\begin{eqnarray}\label{eq:rtkepe}
R&=&a(1-e\cos\eta) \\ \nonumber
t&=&\bigg(\frac{a^3}{G M}\bigg)^{1/2}(\eta - e\sin \eta),
\end{eqnarray}
where $\eta$ is the `eccentric anomaly', 
$a=L^2/[G M (1-e^2)]$ is the semi-major axis, $e=1+[2E L^2/(GM)^2]$ is the orbital eccentricity, $E=1/2V^2-GM/R$ and ${\bf L}={\bf R}\times {\bf V}$ are the specific energy and angular momentum, respectively. 
Thus, Equation~(\ref{eq:rtkepe}) reduces to~(\ref{eq:rtkep}) for $L=0$ (i.e. radial orbits), where $R_{\rm ta}=2a$ and $t_{\rm ta}=\pi(a^3/GM)^{1/2}$.
In general for $L\ne 0$ the maximum separation between the two particles is $R_{\rm ta}=a(1+e)$ at $\eta=\pi$, whereas the closest approach occurs at $\eta=2\pi$ with a pericentric distance $a(1-e)$. 
At any given time the radial and tangential velocity components of the relative trajectory can be written as
\begin{eqnarray}\label{eq:vkep}
V_{\rm rad}&=& \bigg(\frac{a}{G M}\bigg)^{1/2}\frac{e\sin \eta}{1-e\cos\eta}\\ \nonumber
V_{\rm tan}&=& \bigg(\frac{a}{G M}\bigg)^{1/2}\frac{\sqrt{1-e^2}}{1-e\cos\eta}.
\end{eqnarray}

It is clear that under the SCM assumptions the mass appearing in Equation~(\ref{eq:vkep}), henceforth denoted as the {\it timing mass} ($M_{\rm tim}$), necessarily corresponds to the combined virial mass of the two galaxies
\begin{equation}\label{eq:mtim}
M_{\rm tim}=M_{\rm vir,1}+M_{\rm vir,2}.
\end{equation}
This relation holds if (i) galaxies have a finite size $R_{\rm vir}$ and (ii) the separation between the main galaxies is much larger than their combined size, such that $R\gg R_{\rm vir,1}+R_{\rm vir,2}$ at all times.

\subsection{Extra-virial mass}\label{sec:extra}
In the SCM galaxies are modelled as spheres with a finite size $R_{\rm vir}$ and enclosed mass $M_{\rm vir}$. This picture is clearly at odds with the results from cosmological simulations, where haloes exhibit smooth density profiles that extend well beyond the nominal virial radius (see Section~\ref{sec:apostle}). 

Recently, Diemer \& Kravtov (2014; hereafter DK14) show that the mean outermost ($R\gtrsim R_{\rm vir}$) density profile of DM haloes can be fitted by
\begin{eqnarray}\label{eq:dk14}
\rho(r)&=&\rho_{\rm inner}\times f_{\rm trans} +\rho_{\rm bg}\\ \nonumber
\rho_{\rm inner}&\simeq& \rho_{\rm NFW}\equiv\frac{\rho_s}{(r/r_s)(1+r/r_s)^2} \\ \nonumber
f_{\rm trans}&=&\bigg[1+\bigg(\frac{r}{R_{\rm sp}}\bigg)^\beta\bigg]^{-\gamma/\beta} \\
\rho_{\rm bg}&=&\bar{\rho}\bigg[b_e\bigg(\frac{r}{5R_{200}}\bigg)^{-s_e}+1\bigg].
\end{eqnarray}
Profiles with $s_e>0$ converge towards the mean matter density of the Universe, $\bar{\rho}$, in the limit $r\gg R_{200}$, where $R_{200}$ is the radius where the mean density of the halo is 200 times the mean density of the Universe (see \S\ref{sec:vir}). DK14 fit the parameters of the background density ($\rho_{\rm bg}$) to a suite of cosmological $N$-body haloes, finding mean values of $0.5\lesssim s_e\lesssim 1.7$ and $1\lesssim b_e\lesssim 5$ at $z=0$.

 More et al. (2015) shows that the transition between the inner and outer halo profile roughly occurs at the {\it splashback} radius\footnote{ Recently, Mansfield et al. (2016) show that material at first apocentre after infall tends to pile up on an elliptical ``shell'', rather than on a spherical surface. We find that using Mansfield et al. elliptical radius, or the original ``transition'' radius defined in DK14, barely change the mass estimates derived in \S\ref{sec:res}.}, which roughly scales as
\begin{eqnarray}\label{eq:rt}
R_{\rm sp}=0.54[1+0.53 \Omega_m]\big(1 +1.36e^{-\Gamma/3.04}\big)\times R_{200}.
\end{eqnarray}
where
\begin{eqnarray}\label{eq:Gamma}
\Gamma\equiv \frac{\Delta \log_{10}M_{\rm vir}}{\Delta \log_{10} a},
\end{eqnarray}
is the growth rate between the redshifts $z=0.5$ and $z=0$, which correspond to scale factors $a=(1+z)^{-1}=2/3$ and $a=1$, respectively. 
The function $f_{\rm trans}$ steepens the density profile at $R\gtrsim R_{\rm sp}$.   
Note that according to DK14 the slopes $\beta=6$ and $\gamma=4$, such that $\rho\sim r^{-4}$ at $r\gtrsim R_{\rm sp}$. This is a crucial observation as, in contrast to the NFW model, the mass enclosed within the DK14 profile converges asymptotically towards a finite value $M_{\rm out}$ at large radii, where
\begin{eqnarray}\label{eq:Minf}
M_{\rm out} =4\pi\int_0^{R_{\rm vir}} \d r r^2\delta \rho(r)+4\pi\int_{R_{\rm vir}}^{r_{\rm out}} \d r r^2\delta \rho(r)\equiv M_{\rm vir}+\Delta M,
\end{eqnarray}
and $\delta \rho=\rho-\rho_{\rm bg}$.
The integral on the right-hand side of~(\ref{eq:Minf}) can be solved analytically in the limit $r_{\rm out}\to \infty$ under the approximation $\rho_{\rm NFW}\sim r^{-3}$ at $r\gtrsim R_{\rm vir}$, which yields
\begin{eqnarray}\label{eq:DelM}
\Delta M \simeq \frac{1}{4}\frac{M_{\rm vir}}{\ln(1+c_{\rm vir})-c_{\rm vir}/(1+c_{\rm vir})}\bigg(\frac{R_{\rm sp}}{R_{\rm vir}}\bigg)^4 F^1_{2}\bigg[\frac{2}{3},\frac{2}{3},\frac{5}{3},-\bigg(\frac{R_{\rm sp}}{R_{\rm vir}}\bigg)^6\bigg]
\end{eqnarray}
where $F^1_{2}$ is the hyper-geometric function. For convenience, we have replaced the NFW parameters $\rho_s$ and $r_s$ in~(\ref{eq:dk14}) by the virial mass $M_{\rm vir}=4\pi \rho_s r_s^3[\ln(1+c_{\rm vir})-c_{\rm vir}/(1+c_{\rm vir})]$ and the concentration $c_{\rm vir}\equiv R_{\rm vir}/r_s$. 

\begin{figure}
\includegraphics[width=84mm]{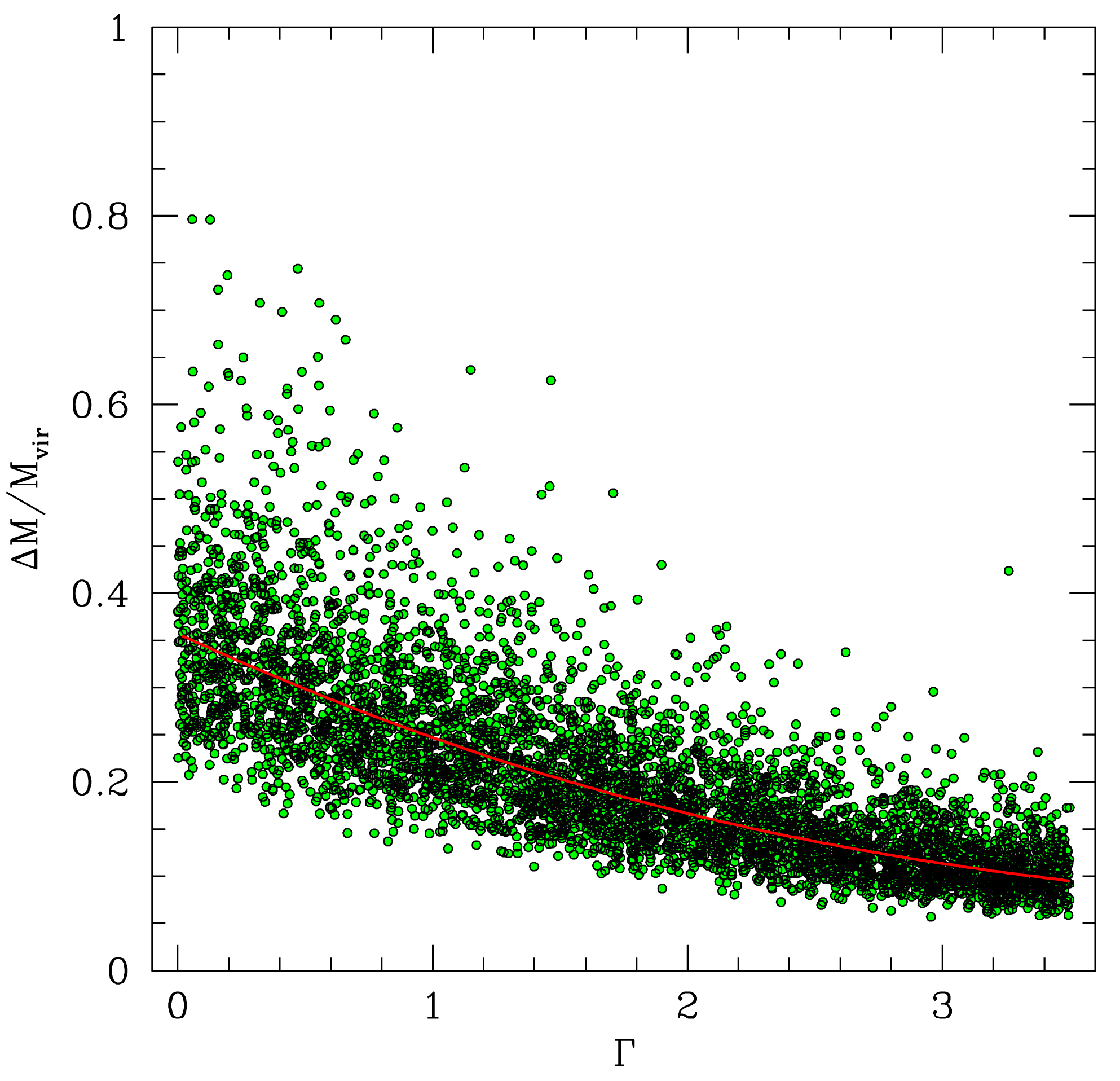}
\caption{Extra-virial mass $\Delta M=M_{\rm out}- M_{\rm vir}$ associated to galaxies with $M_{\rm vir}=10^{12}M_\odot$ and concentration $c_{\rm vir}=9.3$ (red line) as a function of the halo growth rate $\Gamma=\Delta \log_{10}(M_{\rm vir})/\Delta \log_{10} (a)$ calculated between the redshifts $z=0.5$ and $z=0$ (see text). Dots show the effect of introducing a 0.20 dex scatter in the mass-concentration relation. Note that the amount of extra-virial mass is substantial low in galaxies with slow accretion rates ($\Gamma\approx 0$).}
\label{fig:mass_appr}
\end{figure}

To estimate the amount of `extra-virial' mass in Milky Way-sized haloes we show in Fig.~\ref{fig:mass_appr} the value of $\Delta M/M_{\rm vir}$ for a halo with $M_{\rm vir}=10^{12}M_\odot$, $R_{\rm vir}=220\kpc$ and $R_{200}=1.2R_{\rm vir}$ as a function of the growth rate $\Gamma$. The transition radius $R_{\rm sp}$ is given by~(\ref{eq:rt}), whereas the concentration follows from the empirical formula of Mu\~noz-Cuartas et al. (2011), which yields $ c_{\rm vir}=9.31$. For simplicity, we assume that the relatively large scatter in the mass-concentration relation observed in DM-only cosmological simulations follows a Gaussian distribution with a standard deviation of 0.15 dex in $\log_{10}c_{\rm vir}$. Note the substantial amount of extra-virial mass in haloes that experience small accretion rates at $z\lesssim 0.5$.

Adopting this result at face value suggests that the mass perturbing the local Hubble flow, $M_{\rm lhf}$, and the mass derived from the timing argument, $M_{\rm tim}$, may be {\it larger} than the combined virial masses of the galaxy pair, $M_{\rm vir}=M_{\rm vir,1}+M_{\rm vir,2}$, and that the discrepancy will increase in systems with a quiescent mass assembly history at $\mbox{z<0.5}$. We shall return to this issue in \S\ref{sec:res}.

\begin{figure*}
\includegraphics[width=176mm]{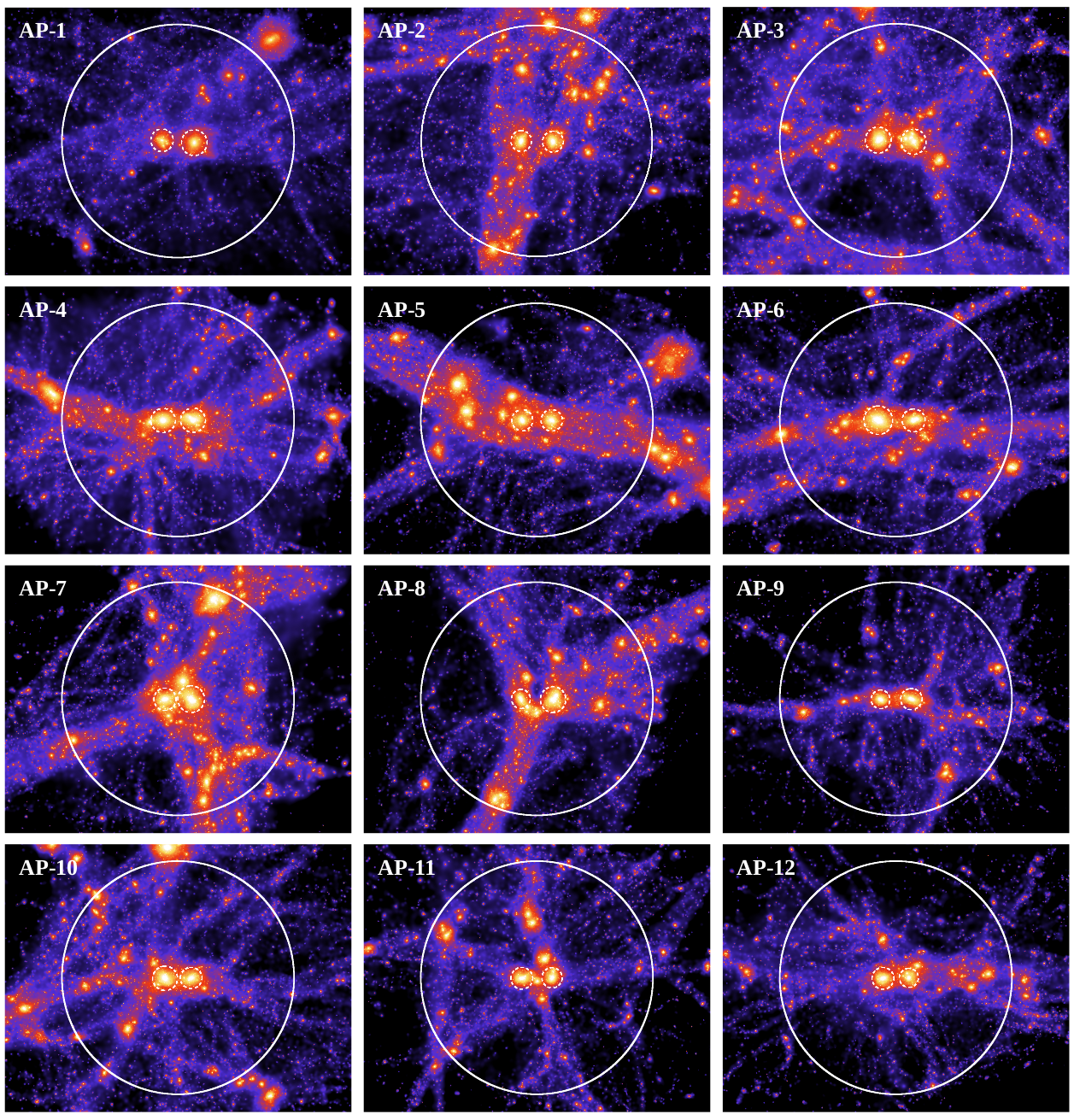}
\caption{Projection of the 12 {\sc Apostle} simulations. Particles are colour-coded according to the local dark matter density. Solid and dashed circles mark a $3\mpc$ radius from the barycentre of the main galaxy pair and their individual virial radii, respectively. }
\label{fig:apostles}
\end{figure*}

\begin{figure*}
\includegraphics[width=176mm]{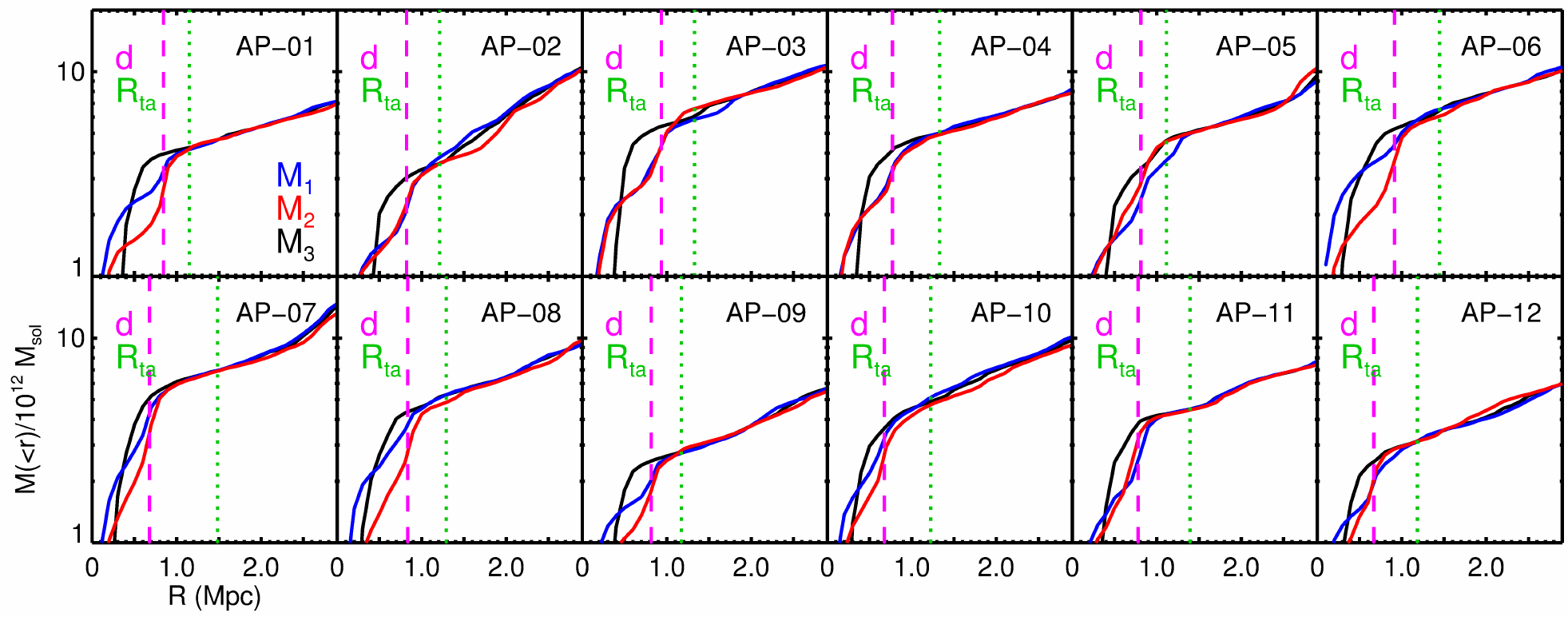}
\caption{Enclosed mass as a function of radius for the 12 {\sc Apostle} simulations. Blue and red lines denote masses measured from the galaxies 1 and 2, respectively, whereas black lines show the enclosed mass measured from the barycentre of the system. Vertical dashed and dotted lines mark the separation between the main galaxies ($d$) and the zero-velocity radius ($R_{\rm ta}$), respectively. Note that the mass profiles grows steeply at $R\lesssim d$, raising monotonically and more gently at $r\gtrsim d$. }
\label{fig:plot_Md_Mr0}
\end{figure*}
\section{Application to Local Group systems}\label{sec:num}
The Spherical Collapse Model (SCM) establishes an exact equivalence between the virial mass of a galaxy and the mass that perturbs the Hubble flow locally.
However, given the highly idealized assumptions on which the SCM model rests it is worth inspecting whether the correspondence $M_{\rm vir}=M_{\rm tim}=M_{\rm lhf}$ holds in realistic systems. To this end we use observations of the Local Group and the {\sc Apostle} $N$-body simulations of Local Group systems (Sawala et al. 2016; Fattahi et al. 2016) as test beds to discuss the accuracy of the three virial mass estimates introduced above.

\subsection{The {\sc Apostle} simulations}\label{sec:apostle}
We construct mock data sets of the local ($<3\mpc$) cosmological
environment using twelve Local Group analogues of the APOSTLE project,
a suite of zoom-in simulations that describe the local environment
(Sawala et al 2016, Fattahi et al. 2016). These Local Group candidates
were selected from the DOVE cosmological $N$-Body simulation described
by Jenkins (2013). A comprehensive discussion on the selection of
Local Group candidates is presented in Fattahi et al. (2016). Here we
bring a summary of the main points.

We use the dark matter-only version of the APOSTLE simulations, at the
resolution level L3 with the dark matter particles mass of
$\sim8.5\times10^{6} M_\odot$. The simulations were performed using
the Tree-PM code P-Gadget3, a private version of the publicly
available code Gadget-2 (Springel et al. 2005). As quoted in table A1
of Fattahi et al. 2016, the Lagrangian regions of the zoom-in
simulations have radii larger than $3\mpc$. Therefore, our mock data
sets at $<3\mpc$ regions are not affected by the boundary effects of
the zoom-in region\footnote{We repeated our analysis using the parent
  simulation, DOVE, to confirm the boundary effects are not important,
  and we found no significant changes in the
  results.}.

Self-bound structures in the simulations are identified using a two step
procedure. First, dark matter halos were picked out using a
friends-of-friends algorithm (FoF; Davis et al 2005) with a linking
length of 0.2 times the mean interparticle separation. Then self-bound
substructures (subhalos) within each FoF halo were found iteratively
using the {\sc subfind} algorithm (Springel 2001). The Local Group
candidates were chosen from pairs of FoF halos, or pairs of massive
subhalos sharing a single FoF halo, whose kinematics are consistent
with observations of the Local Group. The pair members satisfy the
following criteria at z=0

\begin{itemize}
\item separation of 600-1000 $\kpc$,
\item relative radial velocity of -250 to 0 $\kms$,
relative tangential velocity less than 100 $\kms$,
\item sum of the virial masses ($M_{200}$) of the two members in the range $\log(M_{tot}/M_\odot)=[12.2,12.6]$,
\item existence of no subhalo more massive than the smaller of the pair members in $2.5\mpc$ from the barycentre.
\end{itemize}
Table~\ref{tab:nbody} lists some parameters of interest for the twelve
{\sc Apostle} models. Note that the separation between the main {\sc
  Apostle} haloes, d, is in all cases $ R_{\rm vir,1}+R_{\rm
  vir,2}\lesssim d<1\mpc$. All models obey $R_{\rm vir}\ll d$.

APOSTLE adopted the $\Lambda$CDM cosmological parameters according to
WMAP-7, with a matter density of $\Omega_m=0.272$, a dark energy
density of $\Omega_{\Lambda}=0.728$, baryon density of
$\Omega_{b}=0.0455$, Hubble parameter $h=0.704$, linear power spectrum
normalisation $\sigma_8=0.81$, and a spectral index of primordial
power spectrum of $n_s=0.967$. The age of the Universe that results
from the above parameters is $t_0=13.75\gyr$.

Fig.~\ref{fig:apostles} shows a projection of the 12 {\sc Apostle} simulations. Particles are colour-coded according to the local density value. For ease of reference we mark with dashed lines the virial radii of the main galaxies, and with solid lines a $3 \mpc$ radius from the barycentre of the Local Group analogues. This plot highlights the variety in the number, masses and distribution of substructures around Local Group systems. E.g., while some systems are located at the intersection of filaments (e.g. AP-07), other do not show well-defined, large-scale structures in their vicinity (e.g. AP-01). Also, although some {\sc Apostle} models do not have massive substructures in the vicinity of the main haloes (e.g. AP-01 and AP-09), the majority of the Local Group models contain large associations within $3\mpc$ (e.g. Ap-02, AP-05, AP-07, AP-08, AP-10 and AP-11).

It is clear that the SCM fails to capture the complexity of these models. Adopting spherical symmetry, for example, is clearly at odds with the highly anisotropic distribution of haloes around the main galaxies. Also, most Local Group analogues have massive associations at $R\gtrsim 3\mpc$ whose gravitational attraction is neglected by Equation~(\ref{eq:kep}).
Indeed, the SCM states that DM haloes can be modelled as isolated spheres of a finite size $R_{\rm vir}$, thus setting the amount of extra-virial mass to $\Delta M=0$ by construction. To illustrate how poor an approximation this is we plot in Fig.~\ref{fig:plot_Md_Mr0} the mass profile of the {\sc Apostle} systems measured from galaxies 1 and 2 (blue and red solid lines, respectively), as well as from the barycentre of the pair (black solid lines). These profiles show remarkable similarities. In particular we can distinguish two well-defined radial intervals: at small distances $R\lesssim d$, $M(<R)$ is dominated by the individual haloes and raises steeply, whereas at $R\gtrsim d$, the mass profile tends to increase more gently as $M(<R)\sim R^{\alpha}$, with slopes $0.6\lesssim \alpha\lesssim 1.3$, in gross agreement with the background profile given by Equation~(\ref{eq:dk14}). 
At $R\sim d$ the three profiles converge to a single curve. It is worth noting that at the turn-around radius, $R_{\rm ta}/d\sim 1.4$--$2.0$ (see Table{\ref{tab:nbody}), where $R_{\rm ta}$ is given by Equation~(\ref{eq:rta}), the background density already contributes to a significant fraction of the enclosed mass.  
Fig.~\ref{fig:plot_Md_Mr0} confirms that in realistic simulations of structure formation there is no clear-cut physical boundary between virialized structures and the background Universe, as discussed in \S\ref{sec:extra}.

To inspect the accuracy of the SCM in modelling the Hubble flow in the vicinity of the {\sc Apostle} haloes (see also Section~\ref{sec:dis}) we generate mock catalogues of substructures using halos and subhalos more massive than $M_{\rm sub}=10^{9} \,M_{\rm \odot}$, where $M_{\rm sub}$ is the bound mass according to {\sc SUBFIND}. Given the particle resolution of APOSTLE, these objects are resolved with more than 100 particles. The position of each halo/subhalo is defined as the position of the particle with the minimum gravitational potential energy. Velocities correspond to the centre of mass velocities of particles bound to a given halo or subhalo.
To account for the displacement of the centre of mass of the main haloes induced by the presence of massive substructures we compute the location and velocity of the main haloes from the barycentres of the main haloes and its five most massive subhalos at $z = 0$.

In the following Section we study the kinematics of the substructures around the main {\sc Apostle} haloes and outline a Bayesian technique to infer their masses from observations the local Hubble flow.

\begin{figure*}
\includegraphics[width=174mm]{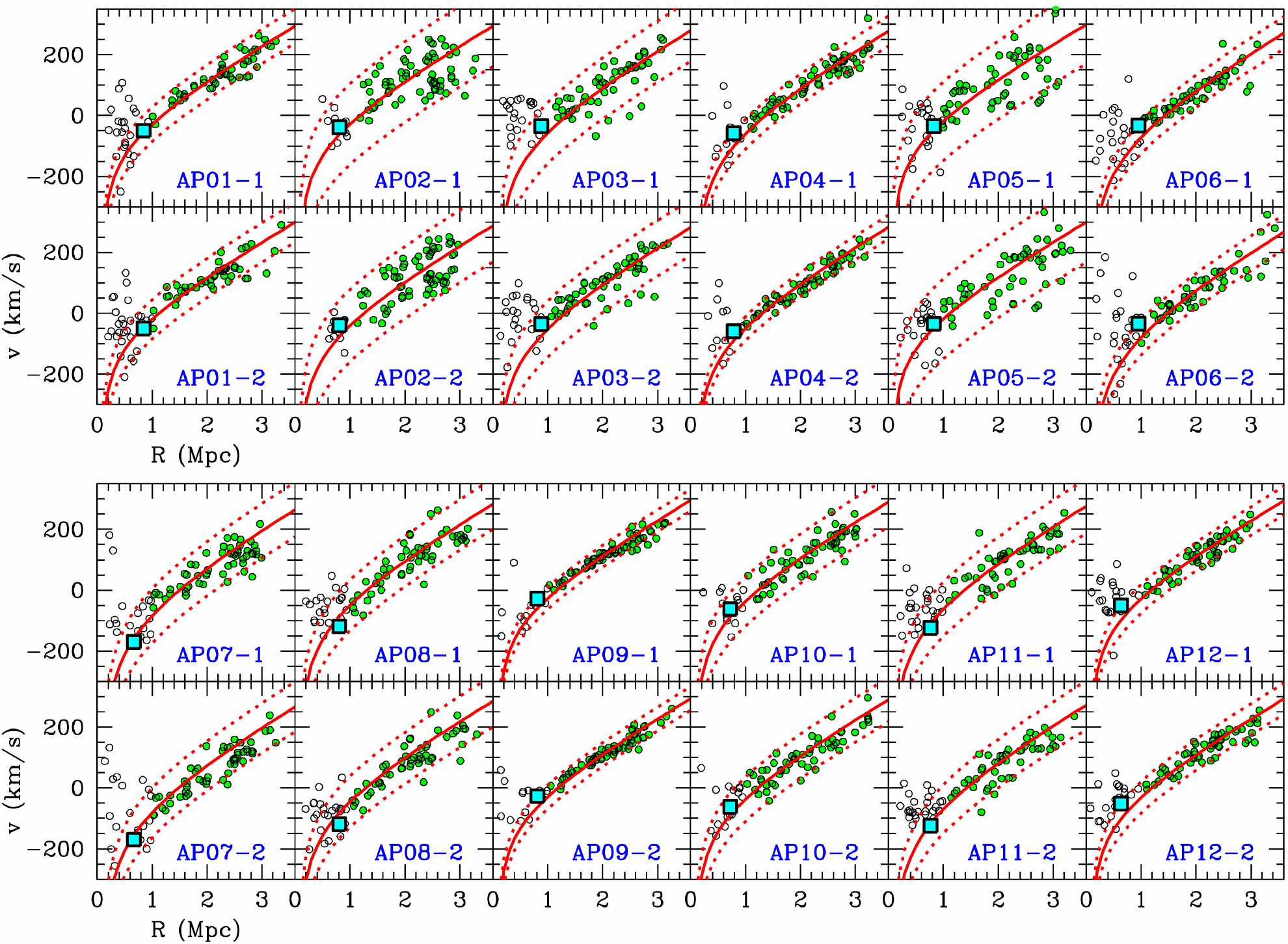}
\caption{Distance velocity relation of DM substructures in the {\sc Apostle} simulations. We fit the distance and radial velocities of 50 random haloes located between $1\mpc$ and $3\mpc$ (green filled dots). Cyan filled squares mark the relative separation and radial velocity of the main galaxies in the {\sc Apostle} models.
Red-solid lines show the best-fitting isochrones $v(R)$ given by Equation~(\ref{eq:vd}). Red-dotted lines show $v(R)\pm 2\sigma_m$, where $\sigma_m$ is the hyperparameter that accounts for the scatter in the distance-velocity relation induced by peculiar motions (see \S\ref{sec:fits}). Notice the remarkable absence of outliers despite of the simplicity of the dynamical models outlined in \S\ref{sec:flow}.}
\label{fig:rv_12}
\end{figure*}
\subsection{Bayesian fits of the perturbed flow}\label{sec:fits}
To measure the mass that perturbs the Hubble flow in the vicinity of the main {\sc Apostle} haloes we fit the distance and radial velocities of sets of $N_g=50$ tracer haloes chosen randomly between $1$--$3\mpc$ from the barycentre of the {\sc Apostle} pairs. This number roughly matches the catalogue size of known galaxies within $\sim 3\mpc$ from the Local Group (see P14 for details). In \S\ref{sec:uncer} we discuss how the number of tracers affect the estimates of the model parameters.
Given that the location of the Local Group barycentre is not an observable quantity we limit the amount of observational constraints to the distance $D$ and radial velocity $v_{h}$ of nearby galaxies that an `observer' would measure from the centre of each of the main haloes. The barycentre location is inferred from the measured distance-velocity relation of tracer substructures by modelling the mass ratio of the main haloes, which implicitely assumes that the main galaxies dominate the mass budget of the Local Group (see also \S\ref{sec:masssub}). For simplicity, we also assume that the observed distances and velocities have no associated errors.

Here we follow Karachentsev \& Makarov (1996) to calculate velocities in the Local Group frame. To simplify our notation, let us assume that the observer sits at the barycentre of galaxy 1. The velocity vector of the observer with respect to the Local Group barycentre (the so-called {\it apex}) can be written as
\begin{equation}\label{eq:apex}
{\bf v}_\odot= - \frac{v_{\rm 2}}{1+f_{\rm m}} \hat {\bf r}_{2};
\end{equation}
where $v_{\rm 2}$ and $\hat {\bf r}_{\rm 2}$ are the radial velocity and the unit vector of galaxy 2, respectively, both measured from the centre of galaxy 1, and $f_{\rm m}=M_{\rm lhf,1}/M_{\rm lhf,2}$ is the mass ratio between the two galaxies (note that the indices 1 and 2 can be exchanged without loss of generality).  
The radial velocity of a tracer galaxy with respect to the LG barycentre can be calculated as
\begin{equation}\label{eq:vrad}
V=v_{h} + \Delta v;
\end{equation}
where $v_h$ is the radial velocity measured by the observer, and $\Delta v$ is the projection of the galaxy position vector onto the solar apex, i.e. $\Delta v={\bf v}_\odot\cdot \hat {\bf r}_g$ and $\hat {\bf r}_g={\bf r}_g/D$, where ${\bf r}_g$ is the position vector of the galaxy measured in the observer frame.

Similarly, the location of the galaxies with respect to the Local Group barycentre can be calculated as
\begin{equation}\label{eq:rlg}
{\bf R}= {\bf r}_g- \frac{d}{1+f_{\rm m}} \hat {\bf r}_{2},
\end{equation}
where $d$ is the separation between the main galaxies.

\begin{figure*}
\includegraphics[width=174mm]{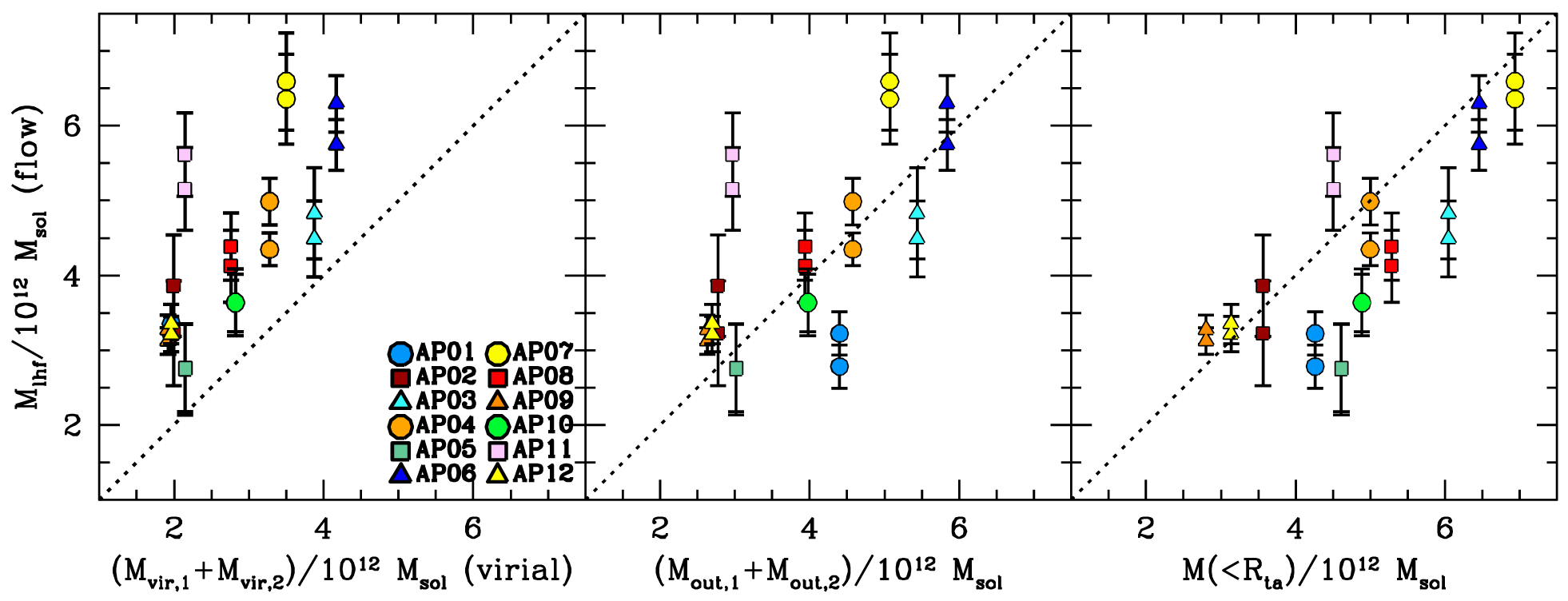}
\caption{Mass derived from the local Hubble flow ($M_{\rm lhf}$) against the {\it combined} virial mass of the main galaxies, $M_{\rm vir,1}+M_{\rm vir,2}$ (left panel), the combined extrapolated mass, $M_{\rm out,1}+M_{\rm out,2}$ (middle panel), and the turn-around radius, $M(<R_{\rm ta})$ (right panel). In this and the following Figures error bars denote 68\% confidence intervals of the posterior distributions of the parameters.}
\label{fig:mdr0}
\end{figure*}

To fit the dynamics of nearby tracer haloes we use a simple Bayesian technique. First, we adopt the following likelihood function (see P14 for a more detailed description)
\begin{equation}\label{eq:likelv}
\mathcal{L}_{\rm LV}(\{D_i,v_{r,i}\}^{N_g}_{i=1}|\vec{S})=\prod_{i=1}^{N_g}\frac{1}{\sqrt{2\pi \sigma_i^2}}\exp\bigg[-\frac{(v_{i}-v_{h,i})^2}{2\sigma_i^2}\bigg];
\end{equation}
where $\vec{S}=(M_{\rm lhf},f_m,\sigma_m)$ is a vector that encompasses the parameters of the model; $D$ and $v_h$ are distances and radial velocities measured from the centre of one of the main galaxies, respectively. For a given set of $\vec{S}$ we use the isochrone~(\ref{eq:vd}) to find the LG-centric velocity
 $V=V(R,M_{\rm lhf},t_0)$, which is then corrected by the apex motion~(\ref{eq:apex}) as $v_{i}=V(R_i,M_{\rm lhf},t_0) - {\bf v}_{\odot} \cdot \hat {\bf r}_{g,i}$ and inserted into the likelihood~(\ref{eq:likelv}). Unless otherwise indicated, all our fits adopt flat priors on $M_{\rm lhf}$, $\log_{10}f_m$ and $\sigma_m$ over ranges that include reasonable parameter values.

Equation~(\ref{eq:kep}) is solved upon the assumption that galaxies move on radial orbits. Tangential motions are accounted for by the hyperparameter $\sigma_m$, which we introduce in the two-dimensional variance of the $i$-th measurement as $\sigma_i^2=\sigma_m^2.$ Tests carried by P14 show that marginalizing over $\sigma_m$ yields unbiased joint bounds on the parameters of interest insofar as the tangential velocities that give rise to the peculiar motions are randomly oriented on the sky seen by the observer.

We apply a nested-sampling technique (Skilling 2004) in order to calculate posterior distributions for our parameters and the evidence of the model. In particular we use the code {\sc MultiNest}, a Bayesian inference tool which also produces posterior samplings and returns error estimates of the evidence (Feroz \& Hobson 2008, 2009). Flat priors are adopted for all model parameters. Table~\ref{tab:fits} lists the median and the $68\%$ standard deviation of the posterior distributions returned by {\sc MultiNest} for the parameters $M_{\rm lhf}$, $f_m$, and $\sigma_m$, together with the derived quantities $M_1$ and $M_2$.

In Fig.~\ref{fig:rv_12} we plot the distance velocity relation of DM substructures in the {\sc Apostle} models as measured from each of the main galaxies\footnote{More precisely, the values of ${\bf R}$ and $V$ plotted here are calculated using the `observed' distance and velocities of the galaxies in the sample plus the Local Group barycentre inferred from the median value of the $f_m$ posterior.}. Green-filled dots correspond to the substructures that we incorporate in the Bayesian analysis. Red-solid lines show the isochrones $V(R)$ associated to the median of the posterior parameters. This plot shows a clear relation between the velocity dispersion of the Hubble flow and the number of massive substructures around the {\sc Apostle} haloes. For example, the models AP-01 and AP-09, which appear fairly isolated and devoid of massive haloes within a $3\mpc$ volume in Fig.~\ref{fig:apostles} exhibit a relatively cold Hubble flow, whereas \mbox{AP-02} and \mbox{AP-05}, with filamentary structures as well as nearby massive associations surrounding them, show clear evidence of large tangential motions, which increase the scatter of the distribution of peculiar velocities.

The good match of Equation~(\ref{eq:vd}) to the local Hubble flow around the {\sc Apostle} haloes is rather remarkable given the simplicity of Equation~(\ref{eq:vd}) and the idealized assumptions on which the SCM rests. Furthermore, we also find that the magnitude of the scatter introduced by peculiar motions in the Hubble flow is comparable to the median value of the hyperparameter $\sigma_m$. Indeed, it is worth noticing the absence of outliers ($|V-V(R)|/\sigma_m>3$) in any of the 24 models\footnote{We repeated our analysis for several suites of $N_g=50$ substructures randomly chosen from the {\sc Apostle} catalogues, but found no outliers beyond a 3$-\sigma$ level in any of the samples.}, which indicates that Equation~(\ref{eq:kep}) provides an {\it effective} description of the dynamics of individual, mass-less tracer particles in the vicinity of local over-densities (see \S\ref{sec:dis} for a discussion). In addition, the relative distance and velocity of the main galaxies (marked with cyan squares) also appears to fall on top of the isochrones~(\ref{eq:vd}) in most models, suggesting a positive correlation between the Local Group mass derived from the classic timing argument ($M_{\rm tim}$) and that obtained from the local Hubble flow ($M_{\rm lhf}$), an issue which \S\ref{sec:tim} inspects in some detail.

\section{Comparison between cosmological masses }\label{sec:res}
In this Section we compare the Local Group mass ($M_{\rm lhf}$) and the individual galaxy masses ($M_1$ and $M_2$) derived from the local Hubble flow against a suite of halo masses measured from the {\sc Apostle} $N$-body simulations.

\subsection{Local Group mass}\label{sec:mlhf}
The lack of well-defined boundaries between virialized haloes and the surrounding background medium complicates the interpretation of the halo masses inferred from the perturbed Hubble flow. Indeed, the left panel of Fig.~\ref{fig:mdr0} shows that the masses derived from the dynamics of nearby galaxies lie systematically above the combined virial mass of the main galaxies ($M_{\rm vir,1}+M_{\rm vir,2}$), which in turn indicates that a large fraction of the Local Group mass that perturbs the Hubble flow is not enclosed within the virialized volume around these systems. As expected from Section~\ref{sec:extra}, the middle panel reveals that within statistical uncertainties the effective mass of the perturber corresponds to the combined, asymptotically-convergent mass of the main haloes, i.e. $M_{\rm lhf}\approx M_{\rm out,1}+M_{\rm out,2}$, where $M_{\rm out}$ is computed from Equation~(\ref{eq:Minf}) after plugging in the individual halo parameters listed in Table~\ref{tab:nbody} and taking the limit $r_{\rm out}\to \infty$. 

To what extent does the background density contribute to the measured value of $M_{\rm lhf}$?
Guided by Fig.~\ref{fig:plot_Md_Mr0}, we plot in the right panel of Fig.~\ref{fig:mdr0} the Local Group mass inferred from the perturbed Hubble flow against the mass enclosed within the turn-around radius $M(<R_{\rm ta})$, where $R_{\rm ta}/d\sim 1.4$--$2.0$ (see Table{\ref{tab:nbody}). Comparison of the two mass estimates reveals that $M(<R_{\rm ta})$ lies systematically above $M_{\rm lhf}$, suggesting that at leading order the peculiar motions of nearby ($<3\mpc$) galaxies are governed by the {\it total} mass of the main galaxies, and that the contribution of the local background density to the inferred Local Group mass is not statistically significant.


\subsection{Mass ratio of the main galaxies}\label{sec:fm}
In Section~\ref{sec:fits} we discussed the necessity to model the mass ratio between the main galaxies ($f_m$) in order to convert `observed' distances and velocities of nearby galaxies into Local Group-centric quantities. P14 use geometrical arguments to demonstrate that choosing the wrong mass ratio systematically increases the velocity scatter of the observed Hubble flow (see their Appendix B).

Fig.~\ref{fig:q} shows 
that the values of $f_m$ inferred from the Hubble flow match within statistical uncertainties the virial mass ratios of the main galaxies, and that none of the {\sc Apostle} haloes deviate from the one-to-one correlation (marked with dotted lines for ease of reference) beyond a $2\sigma$-level. Note also the reciprocal relation between the values of $f_m$ inferred from each of the {\sc Apostle} haloes, such that $f_{m,1}\approx f^{-1}_{m,2}$. This is indeed to be expected given that the parameter $f_m$ is defined locally as the mass ratio between the observer's and the neighbour galaxy. For example, the virial mass of the AP-08 haloes (plotted with red squares in Fig.~\ref{fig:q}) are 2.0$\times 10^{12}M_\odot$ and 0.75$\times 10^{12}M_\odot$, respectively (see Table~\ref{tab:nbody}), whereas the mass ratio inferred from these haloes is $2.1\pm 0.5$ and $0.4\pm 0.1$ (see Table~\ref{tab:fits}).

\begin{figure}
\includegraphics[width=84mm]{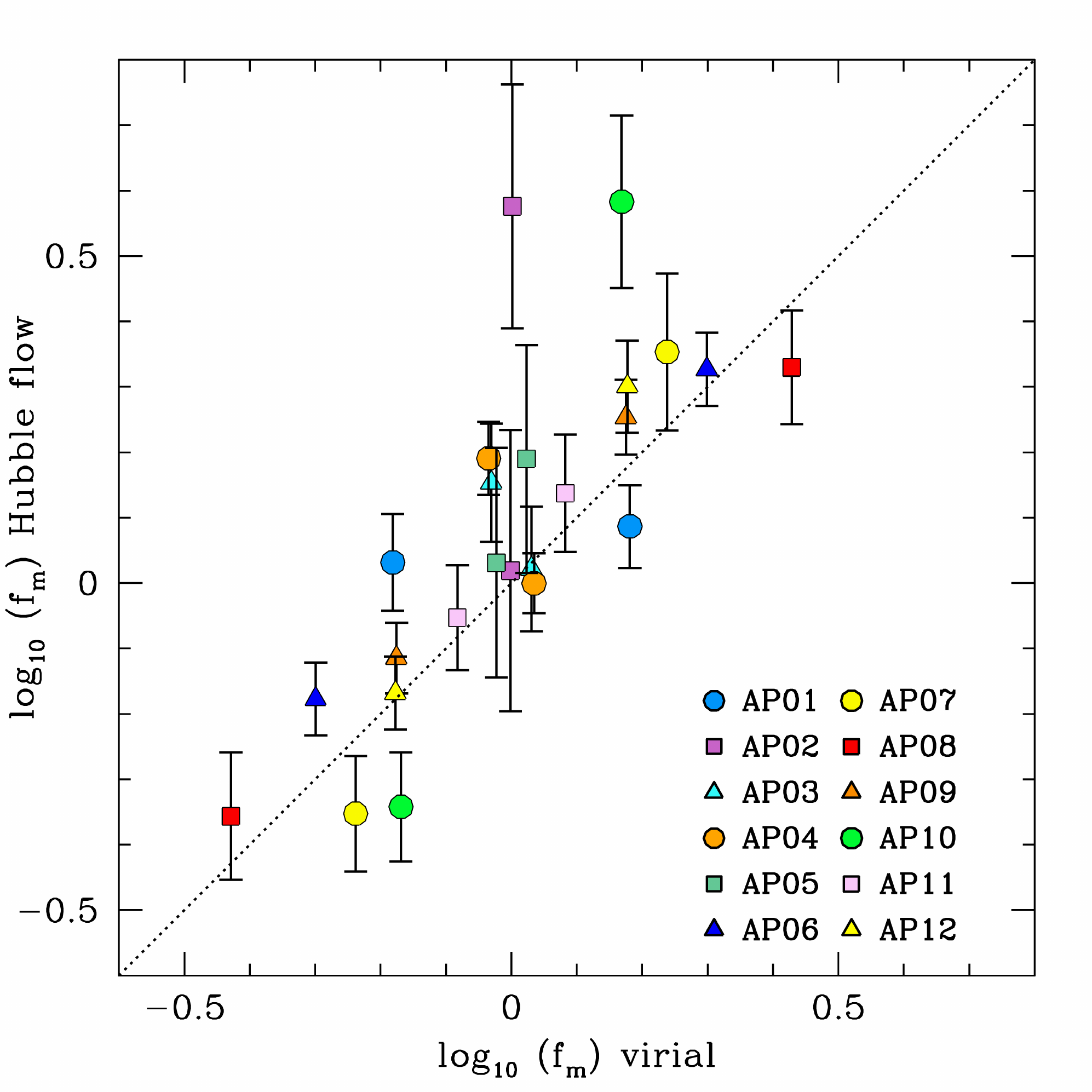}
\caption{Comparison between the mass ratio of the main haloes inferred from the Hubble flow against the virial mass ratio.}
\label{fig:q}
\end{figure}

\subsection{Individual halo masses}
We can now combine the values of $M_{\rm lhf}$ and $f_m$ derived in Sections~\ref{sec:mlhf} and~\ref{sec:fm}, respectively, in order to estimate the individual masses of the main haloes. The underlying assumption here is that the mass inferred from the Hubble flow corresponds to the combined mass of the main haloes, which is a key ingredient in the dynamical models outlined in Section~\ref{sec:fits}. Before discussing the result it is worth noting that the Bayesian fits return the mass of the galaxy where the observer is located plus that of the neighbour galaxy, which yields four individual mass estimates per {\sc Apostle} model and 48 mass estimates in total.

In Fig.~\ref{fig:gamma3} we plot the individual galaxy masses inferred from the Hubble flow ($M$) and compare them against the corresponding virial masses (left panel) and the total mass extrapolated from the DK14 profile (middle panel). As expected, we find that the individual mass derived from the Hubble flow lies systematically above the virial mass. The difference appears to be mildly correlated with the halo accretion rate ($\Gamma$), as suggested by Fig.~\ref{fig:mass_appr}. In particular, haloes with a slow accretion rate tend to contain a larger amount of extra-virial mass with respect with those that experienced recent accretion events, although the relatively small number of {\sc Apostle} haloes (24) combined with the relatively large model uncertainties introduces condiderable scatter in the correlation.

The middle panel of this Figure shows that the asymptotic mass $M_{\rm out}$ derived from the DK14 profile extrapolation (see \S\ref{sec:extra}) 
 accounts for the extra-virial mass and recovers within statistical uncertainties the individual halo masses derived from the local flow. This can be clearly seen in the right panel, where we plot the distribution of the mass residuals, $\Delta M/M_{\rm gal}\equiv (M-M_{\rm gal})/M_{\rm gal}$, where $M_{\rm gal}$ denotes either $M_{\rm vir}$ (red dashed line) or $M_{\rm out}$ (solid blue lines). The median and standard deviations of the distributions are $\langle \Delta M/M_{\rm vir}\rangle =0.54\pm 0.50$, and $\langle \Delta M/M_{\rm out}\rangle =0.09\pm 0.33$. 
Hence, while the individual galaxy masses derived from the Hubble flow ($M$) tend to overe-estimate the corresponding virial masses ($M_{\rm vir}$), they provide an unbiased constraint of the asymptotically-convergent {\it total} masses of the main galaxies ($M_{\rm out}$) derived from Equation~(\ref{eq:Minf}) in the limit $r_{\rm out}\to \infty$.

\begin{figure*}
\includegraphics[width=174mm]{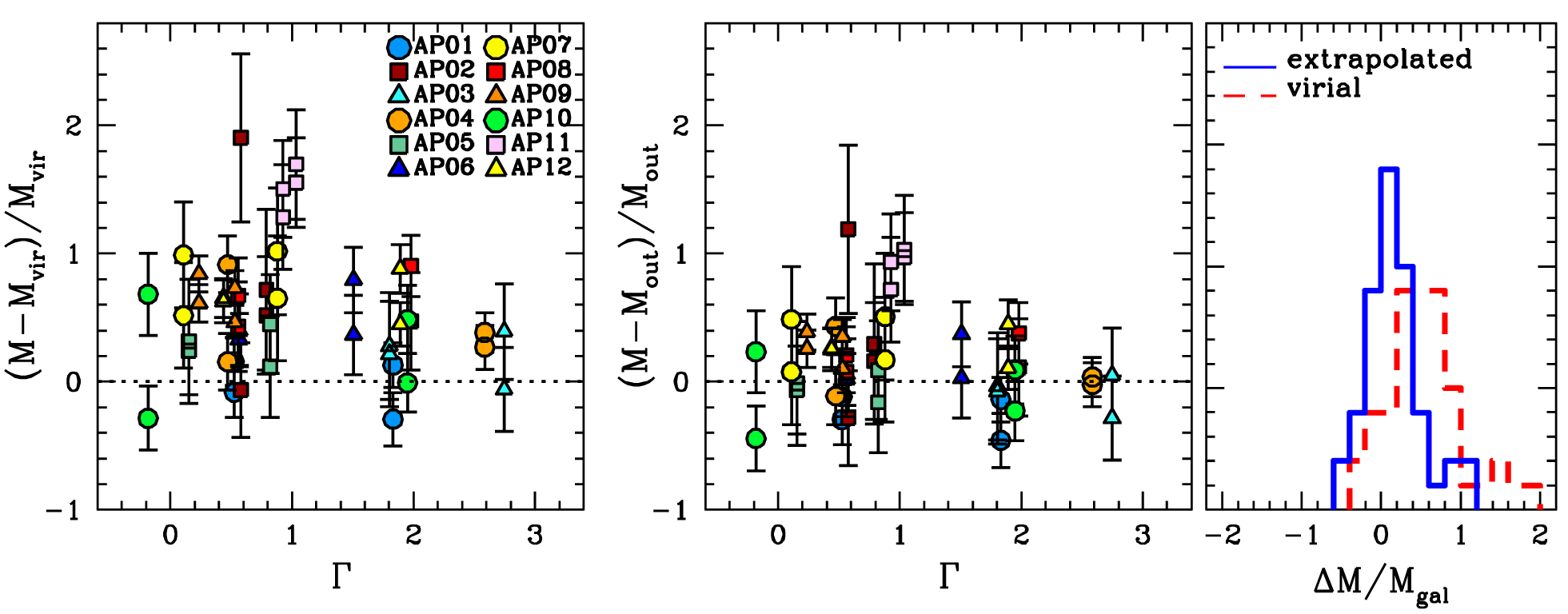}
\caption{{\it Individual} galaxy masses derived from the Hubble flow ($M$) versus the virial mass ($M_{\rm vir}$, left panel) and extrapolated total mass ($M_{\rm out}$, middle panel) 
as a function of the halo growth rate given by Equation~(\ref{eq:Gamma}). Interestingly, the left panel shows a negative correlation between $(M-M_{\rm vir})/M_{\rm vir}$ and $\Gamma$ which roughly follows the trend shown in Fig.~\ref{fig:mass_appr}. The right panel shows a histogram of the residual fraction $\Delta M/M_{\rm gal}\equiv (M-M_{\rm gal})/M_{\rm gal}$, where $M_{\rm gal}$ denotes either $M_{\rm vir}$ (red dashed line) or $M_{\rm out}$ (blue solid line).  Note that within statistical uncertainties the individual halo masses estimated from the local Hubble flow are consistent with the extrapolated masses of the DK14 profile in the limit $r_{\rm out}\to \infty$. }
\label{fig:gamma3}
\end{figure*}

\subsection{The classical timing argument}\label{sec:tim}
The relative motion between the main galaxies of the Local Group has been widely used to constrain the combined mass of the system, as discussed in Section~\ref{sec:timing}. This technique, which is usually known as the `timing argument', provides an estimate of the Local Group mass that is complementary to that derived from the Hubble flow. 

In this work we assume that the relative distance and velocity vector of the companion galaxy are known exactly, which allow us to compute the combined Local Group mass directly from Equation~(\ref{eq:vkep}). Thus, in contrast to the mass estimates derived from the perturbed Hubble flow, which have associated uncertainties owing to the unknown tangential velocities of the tracer haloes, the timing mass has no associated statistical error. Deviations between the timing and the virial mass can be therefore traced back to the idealized SCM assumptions on which Equation~(\ref{eq:vkep}) is based. Note also that the classical timing argument does {\it not} inform on the mass ratio of the main galaxies.

The upper panel of Fig.~\ref{fig:mt} plots the timing argument mass $M_{\rm tim}$ against the combined virial mass of the main haloes in the 12 {\sc Apostle} simulations. The two mass estimates are somewhat correlated, although the scatter is very large. Notice that for approximately half of the halo pairs in our sample the timing mass clearly over-estimates the combined virial mass, in agreement with the recent analysis of McLeod et al. (2016), who use a larger sample of $N$-body models to calibrate the correpondence between $M_{\rm tim}$ and the combined virial mass $M_{\rm vir,1}+M_{\rm vir,2}$. 

In the lower panel we compare the timing argument mass and that derived from the Hubble flow, which also exhibits a positive, strongly scattered correlation. Comparison between the upper and lower panels show that $M_{\rm lhf}\gtrsim M_{\rm tim}$ for the haloes where $M_{\rm tim}\approx M_{\rm vir,1}+M_{\rm vir,2}$, as expected from Section~\ref{sec:mlhf}, although some halo pairs, like AP-08, do not obey this relation. 

The discrepancy between $M_{\rm lhf}$ and $M_{\rm tim}$ can be easily spotted in the Hubble diagram\footnote{Note that Fig.~\ref{fig:rv_12} neglects the contribution of the tangential velocity to the timing argument, which in general tends to boost $M_{\rm tim}$. However, the {\sc Apostle} haloes were chosen to have relatively small tangential motions, as outlined in \S\ref{sec:apostle}.} of nearby galaxies plotted in Fig.~\ref{fig:rv_12}. For example, the relative distance and velocity of the main haloes of AP-06 (cyan dots) lies above the best-fitting isochrone $v(R)$ (red lines), whereas the opposite is true for AP-08, suggesting that the timing masses of AP-06 and AP-08 should under- and over-estimate the value of $M_{\rm lhf}$, respectively, which is indeed visible in Fig.~\ref{fig:mt}.

Our results indicate that on average the mass estimated from the timing argument lies in between the combined virial mass of the main galaxies and the mass perturbing the local Hubble flow. Unfortunately, the limited statistics provided by the twelve {\sc Apostle} pairs prevents a more careful analysis of the dynamical processes that drive the underlying correlation between the three mass estimates. 

\begin{figure}
\includegraphics[width=82mm]{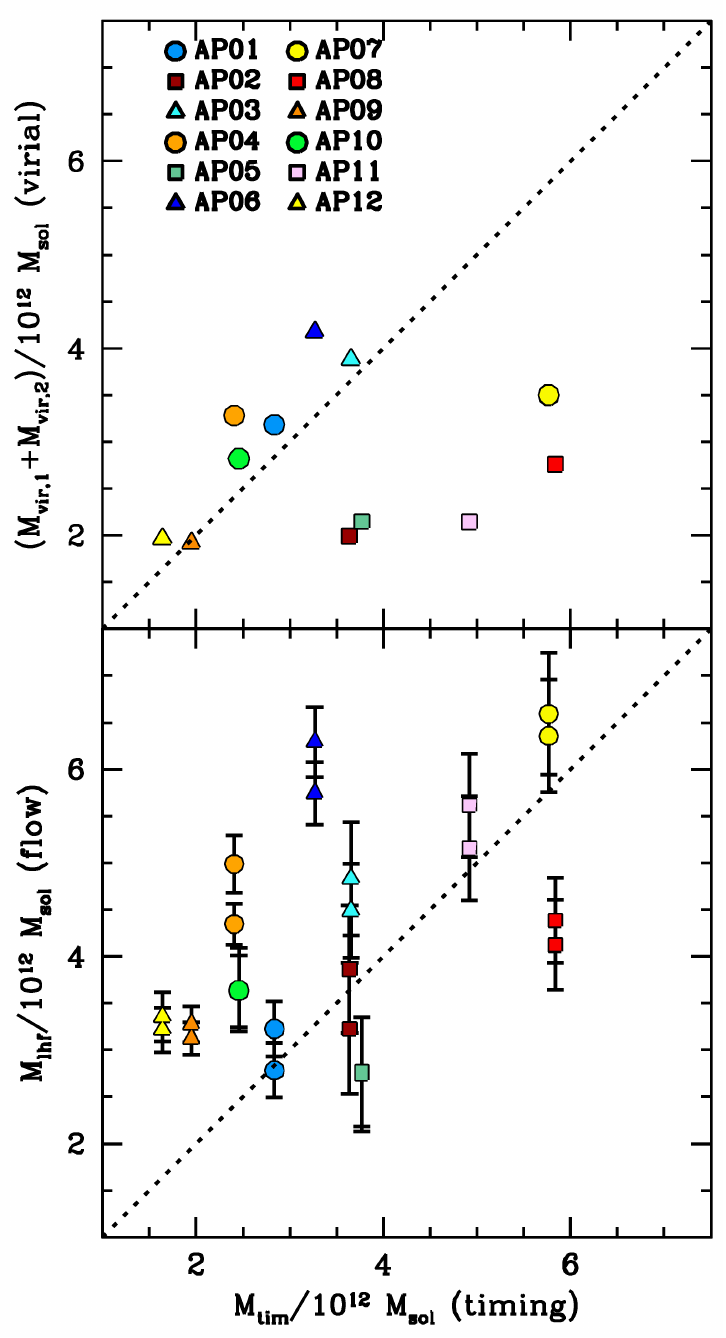}
\caption{Mass derived from the timing mass against the combined virial mass of the main haloes (top panel) and the mass inferred from the local Hubble flow (middle panel). Note that the masses show positive, but strongly scattered correlations.}
\label{fig:mt}
\end{figure}
\section{Discussion}\label{sec:discussion}

\subsection{Model assumptions}\label{sec:dis}
Although the SCM described in Section~\ref{sec:mass} bears little resemblance to the mass distribution found in the {\sc Apostle} simulations, it provides an accurate representation of the peculiar velocities of nearby tracer galaxies (see Fig.~\ref{fig:rv_12}). Such a remarkable --and somewhat unexpected-- agreement is worth investigating in some depth. To this aim we discuss below the validity of some of the key assumptions on which the SCM rests within the Local Group context. Our conclusion is that observations of the local Hubble flow at $R\gtrsim d$, where $d$ is the current separation between the main galaxies, are scarcely sensitive to the detailed mass distribution of the perturber, $M(<d)$, or to its mass assembly history, $M(t)$. 

\subsubsection{Galaxy pair}\label{sec:pair}
Notice first that while Equation~(\ref{eq:kep}) accounts for perturbations in the Hubble flow induced by a {\it single} spherical over-density with radius $R_{\rm vir}$ and mass $M_{\rm vir}$, the Local Group is dominated by {\it two} similar-size galaxies, the Milky Way and Andromeda. Lynden-Bell (1981) argues that the SCM can be still used to describe the dynamics of nearby galaxies upon the adoption of coordinates centred at the barycentre of the Local Group, such that
\begin{eqnarray}\label{eq:angmom}
0&=&M_\g{\bf R}_{\g}+ M_\a{\bf R}_{\a}\\ \nonumber
0&=&M_\g{\bf V}_{\g}+ M_\a{\bf V}_{\a}.
\end{eqnarray}
In this frame galaxies at distances $R\gg d\equiv |{\bf R}_{\rm A}-{\bf R}_{\rm G}|$ feel a potential that can be approximated as 
\begin{eqnarray}\label{eq:potexp}
\Phi(R,\theta)\approx -\frac{G M }{R} + \frac{1}{2}H^2\Omega_{\Lambda}R^2 +\frac{GM f_{\rm m}}{2(1+f_{\rm m})^2}\frac{(1-3 \cos^2\theta)d^2}{R^3};
\end{eqnarray}
where $\cos\theta\equiv \hat{\bf R}\cdot \hat{\bf d}$ and $f_m\equiv M_\g/M_\a$. 
The Keplerian perturbation~(\ref{eq:kep}) is thus recovered by neglecting the right-hand side of Equation~(\ref{eq:potexp}), which 
corresponds to a quadrupole that decays as $\sim (d/R)^3$. 

Tests against controlled $N$-body experiments carried by P14 show that, although the potential quadrupole affects the trajectories of galaxies at $R\lesssim d$, its impact on the local Hubble flow is negligible at $R\gtrsim d$.
Table~\ref{tab:nbody} shows that the main galaxies in the {\sc Apostle} models have $d\lesssim 1\mpc$ without exception, while the mock catalogues generated in \S\ref{sec:fits}
only include galaxies between $1$--$3\mpc$. By design, our lower distance cut removes most of the sensitivity of Equation~(\ref{eq:vd}) to the potential quadrupole generated by the main galaxy pair, as indicated by P14 tests and confirmed in \S\ref{sec:res}.


\subsubsection{Accretion and time-dependence of the potential}\label{sec:time}
The SCM also neglects the effects of mass accretion, effectively assuming that the mass $M$ is constant throughout the expansion and subsequent collapse of the initial over-density. This is at odds with observations of the outskirts of the Milky Way and M31, which reveal clear evidence of past merger events (e.g. see Belokurov et al. 2006; McConnachie et al. 2009).

P14 use dynamical invariants (see Pe\~narrubia 2013 for details) to derive a first-order correction to the perturbed Hubble flow~(\ref{eq:vd}) around time-evolving haloes. For systems with a slow growth rate, $M(t)\approx M_0[1+\epsilon (t-t_0)/t_0]$, the perturbed Hubble flow can be written as
\begin{equation}\label{eq:vdt}
V(R)\approx \bigg[(1.2 + 0.16 \Omega_\Lambda)\frac{R}{t_0} - 1.1\bigg(\frac{GM_0}{R}\bigg)^{1/2}\bigg]\bigg[1+\epsilon t_0\bigg(\frac{GM_0}{R^3}\bigg)^{1/2}\bigg],
\end{equation}
where $\epsilon\equiv (\d M/\d t)(t_0/M_0)$ is the dimension-less growth rate and $M_0=M(t_0)$. If the mass evolution is slow one can related this quantity to the mass accretion rate~(\ref{eq:Gamma}) as $\epsilon\simeq \Gamma \Delta\log_{10}a \ln (10) t_0/\Delta t\approx 1.1 ~\Gamma$ for $\epsilon\ll 1$, where $\Delta t$ and $\Delta\log_{10}a$ are calculated between redshifts $z=0.5$ and $z=0$.

Note that the first-order correction term on the right-hand side of~(\ref{eq:vdt}) declines at large distances from the galaxy as $\Delta V\sim R^{-3/2}$. For systems that accrete mass recently ($\epsilon >0$) the mass evolution steepens the isochrone $V(R)$ at small distances ($R\lesssim d$) from the matter source, and leaves the perturbed Hubble flow approximately invariant at $R\gtrsim d$. Hence, a time-dependent potential and the Local Group quadrupole influence the dynamics of galaxy tracers within a similar distance range, as discussed in \S\ref{sec:pair}. 
The mock data catalogues constructed in Section~\ref{sec:apostle} minimize the impact of both, the hierarchical mass accretion of the main haloes and the potential quadrupole by selecting tracer haloes at distances $R\gtrsim d$ from the Local Group barycentre. Under this particular choice the assumption $M=M_0={\rm const.}$ in Equation~(\ref{eq:kep}) does not bias mass mass inferred from the observed Hubble flow.

\subsubsection{Large Scale Structure}\label{sec:lss}
At distances $R\gg d$ the Hubble flow around the Local Group exhibits pertubations induced by nearby groups of galaxies (e.g. Mohayaee \& Tully 2005; Karachentsev et al. 2009; Courtois et al. 2012) which are not accounted for by Equation~(\ref{eq:kep}). 

A simple remedy for suppressing the impact of external systems on the virial mass estimates is to exclude galaxies in the vicinity of major perturbers. For example, P14 impose a distance cut to the data set at $R_{\rm max}=3 \mpc$, which roughly corresponds to the distance to the closest associations -- Centaurus A, M81 and IC 342 -- and only fit galaxies whose distance to any of the three major associations is larger than $1 \mpc$. Such hard cuts are motivated by the decreasing accuracy of Equation~(\ref{eq:kep}) at distances where the contribution of the Local Group to the local gravitational acceleration becomes negligible. Here we fit the dynamics of tracer substructures within $3\mpc$ from the main haloes regardless of the distribution of massive haloes beyond this volume. Recall that by design the {\sc Apostle} groups do not contain massive neighbours within $\sim 2.5\mpc$ (see \S\ref{sec:apostle}).

Furthermore, the SCM also neglects the overall motion of the Local Group towards peak-density regions in the local volume, such as the Big Attractor (Lynden-Bell et al. 1988), or the vast `wall' of structures connected to the Virgo Cluster (Tully \& Fisher 1987). Numerical simulations of Local Group-like systems show that Large Scale Structures may alter the trajectories of nearby galaxies and induce anisotropic peculiar velocities in the local Hubble flow (Libeskind et al. 2011; Ben{\'{\i}}tez-Llambay et al. 2013). However, recent attempts to detect kinematic anisotropies in the Hubble flow observed within $1$--$3\mpc$ have not yielded any statistically-meaningful deviation from isotropic models (see P16), which suggests that the effects of Large Scale Structure on 
Equation~(\ref{eq:vd}) may be subdominant with respect to those of the Local Group. In essence, the results of P16 indicate that the galaxies within a $3\mpc$ volume around the Local Group exhibit a coherent motion with respect to the surrounding cosmic web.

Tests with {\sc Apostle} simulations support this conclusion and indicate that large scale structures do not unduly affect the dynamical constraints on the Local Group mass derived from the motion of nearby ($R\lesssim 3 \mpc$) galaxies, which greatly simplifies the theoretical framework of \S\ref{sec:mass}.

\subsubsection{Massive substructures}\label{sec:masssub}
The presence of massive satellites in our Galaxy and Andromeda adds uncertainty into the determination of $({\bf R}_\g$, ${\bf V}_\g)$, and $({\bf R}_\a,{\bf V}_\a)$, respectively, which in turn propagates to the location and the motion of the Local Group barycentre (e.g. G\'omez et al. 2015). 

P16 incorporates the mass contribution of the largest satellites of the Local Group (the LMC and M33) by fitting $M$ and the ratios $f_m=M_\g/M_\a$, $f_c=M_\lmc/M_{\rm MW}$ and $f_{M33}=M_{\rm M33}/M_{\rm M31}$ simultaneously, where 
\begin{eqnarray}\label{eq:angmom2}
M_\g{\bf V}_{\g}&=&M_\mw{\bf v}_{\mw}+ M_\lmc{\bf v}_{\lmc}\\ \nonumber
M_\a{\bf V}_{\a}&=&M_{\rm M31}{\bf v}_{\rm M31}+ M_{\rm M33}{\bf v}_{\rm M33},
\end{eqnarray}
and similarly for ${\bf R}_\g$ and ${\bf R}_\a$.

However, the physical interpretation of the satellite masses derived from Equation~(\ref{eq:angmom2}) is complicated by the fact that satellites orbiting a larger system tend to lose their dark matter envelopes to tides after a few pericentric passages (e.g. Pe\~narrubia et al. 2008; 2009). Tidal stripping thus calls for a more sophisticated dynamical modelling of the host-satellite interaction beyond the point-mass approximation on which Equation~(\ref{eq:angmom2}) rests.

In this contribution we account for the presence of substructures within the main galaxy haloes by computing the combined barycentre of the main halo and its five most massive subhalos at $z=0$ (see \S\ref{sec:apostle}). In addition, we construct mocks that neglect the effects of substructures, finding similar bounds on the model parameters. Indeed, inspection of the {\sc Apostle} models shows that the most massive subhaloes tend to be located in the outskirts of galactic haloes, where their relative velocity with respect to the parent halo, and thus their contribution to Equation~(\ref{eq:angmom2}), is small. A more detailed statistical analysis of cosmological $N$-body models by Deason et al. (2014) shows that the presence of satellites as large as the LMC within the inner-most regions of Milky Way-like haloes is typically rare, in agreement with our findings.


\subsection{Model uncertainties \& sample size}\label{sec:uncer}
Although the current census of galaxies in the vicinity of the Local Group ($0.8\lesssim R/\mpc \lesssim 3$) is still relatively small ($N_g\approx 35$ according to P14), on-going observational efforts may soon uncover a large population of faint galaxies predicted by $\Lambda$CDM models (e.g. using HI surveys, Tollerud et al. 2015). In this Section we explore to what degree the sample size affects the constraints on our model parameters. For illustration purposes we use the {\sc Apostle} halo with the largest number of identified dark matter structures (AP11-330892, see Table~\ref{tab:nbody} and Fig.~\ref{fig:apostles}) to generate mock catalogues with a varying sample size of subhaloes within $1$--$3\mpc$ from the barycentre of the AP-07 system. 

Fig.~\ref{fig:number} shows the dependence of the posterior distributions of the parameters $M_{\rm lhf}$ (upper panel), $f_{m}$ (middle panel) and $\sigma_m$ (lower panel) as a function of the number of galaxies in the sample, $N_g$. Error bars indicate 68\% confidence intervals around the median value of the posterior distribution. As expected, the uncertainties of the model parameters tend to decrease for increasing values of $N_g$. All samples produce consistent bounds on $M_{\rm lhf}$ and $f_{m}$ given the computed uncertainties. Interestingly, having large samples is specially important for measuring the mass ratio of the main galaxies $f_m$ with accuracy, whereas the effective mass $M_{\rm lhf}$ appears less sensitive to the value of $N_g$.

The parameter $\sigma_m$, however, does not behave well. 
In particular, the bottom panel of Fig.~\ref{fig:number} shows a sudden increase in the value of $\sigma_m$ at $N_g\gtrsim 100$ which cannot be accounted by statistical uncertainties of the fits. The reason behind the discontinuous dependence of $\sigma_m$ with $N_g$ can be traced back to the presence of long tails in the distribution of peculiar velocities, which only reveal themselves when the number of galaxies in the sample is large enough. 
As discussed in \S\ref{sec:fits}, the uncertainties of $M_{\rm lhf}$ and $f_{m}$ are directly related with the hyperparameter $\sigma_m$, which itself provides a proxy for the scatter of the Hubble flow within the distance interval were the tracers are located. As a result, we find that the size of the error bars derived from galaxy samples with $N_g\lesssim 100$ members tends to be underestimated. 

We have checked that AP-07 is the only {\sc Apostle} model in which $\sigma_m$, and therefore the quoted model uncertainties, have a discontinuous dependence on $N_g$. Yet, given the relatively small number of known galaxies within $3\mpc$ of the Local Group, this result calls for caution when interpreting P14 and P16 statistical bounds on the masses of Local Group galaxies. In particular, the value of $\sigma_m\simeq 50\kms$ recently measured using $N_g\simeq 35$ tracer galaxies within $3.0\mpc$ from the Local Group (P14, Banik \& Zhao 2016) must be taken as a lower limit.

\begin{figure}
\includegraphics[width=82mm]{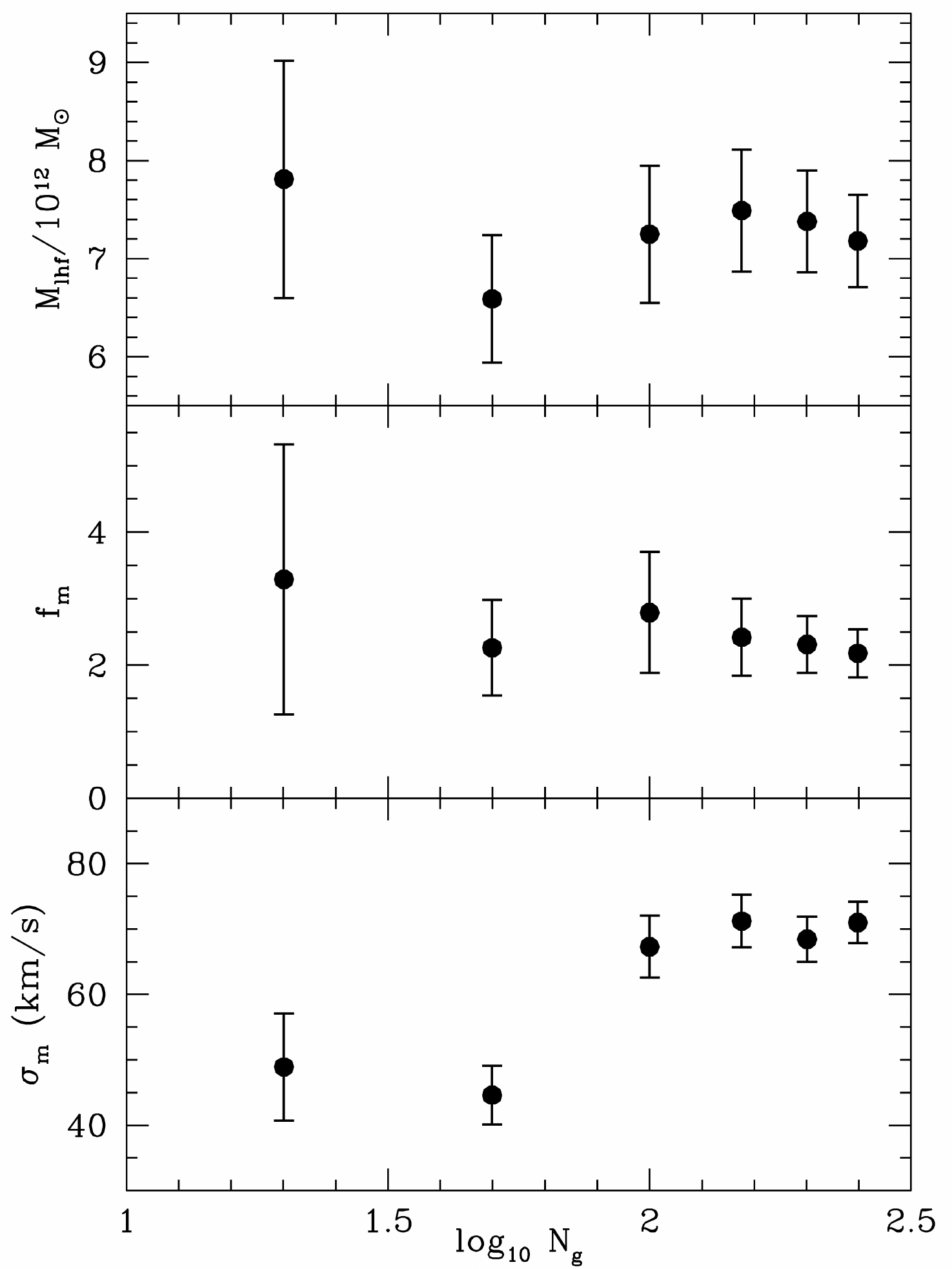}
\caption{Median and 68\% confidence intervals of the posterior distributions of the parameters $M_{\rm lhf}$ (upper panel), $f_{m}$ (middle panel) and $\sigma_m$ (lower panel) of the model {\sc Apostle} halo AP11-330892 (Table~\ref{tab:nbody}) as a function of the number of halo tracers in the mock catalogues. }
\label{fig:number}
\end{figure}
\subsection{The virial mass of the Milky Way}
We now turn to the thorny task of estimating the virial mass of a galaxy, $M_{\rm vir}$, from the effective mass perturbing the local Hubble flow, $M_{\rm lhf}$. Note first that while the parameter $M_{\rm lhf}$ can be measured directly from the heliocentric distances and velocities of nearby galaxies, a derivation of the virial mass of the Milky Way ( $M_{\rm MW,vir}$) requires a priori information on the matter distribution of the dark matter halo. Neglecting the baryonic mass and adopting DK14 profile for simplicity reduces the number of free parameters to three: $M_{\rm MW,vir}$, $c_{\rm vir}$ and $\Gamma$; for one single constraint, i.e. the Milky Way mass inferred from the Hubble flow, $M_{\rm MW,lhf}$. Using the statistical correlation between virial mass and concentration $c_{\rm vir}=c_{\rm vir}(M_{\rm vir},z$) at $z=0$ found in cosmological $N$-body simulations (Mu\~noz-Cuartas et al. 2011) removes one further degree of freedom and reduces the problem to estimating the growth rate of our Galaxy, $\Gamma$. 

As discussed in the Introduction, the virial mass of field haloes grows via (i) the hierarchical accretion of substructures and, (ii) a monotonic decrease of the density threshold with respect to which virial over-densities are defined, which leads to the so-called {\it pseudo-evolution} of dark matter haloes. 
For haloes with a relatively small growth rate Equation~(\ref{eq:Gamma}) can be linearized as
$$\Gamma\approx \Gamma_{\rm acc}+\Gamma_{\rm pseudo}.$$
Diemer et al. (2013) and Zemp (2014) find that in Milky Way-sized haloes pseudo-evolution accounts for $\sim 30\%$ of the virial mass growth from $z=0.5$ to the present, which yields $\Gamma_{\rm pseudo}\sim 0.6$. On the other hand, in our Galaxy $\Gamma_{\rm acc}$ may be dominated by the (total) mass of the brightest satellite galaxy, the Large Magellanic Cloud (LMC), which to date remains strongly debated in the literature.

Let us inspect two leading scenarios, one where the LMC contribution to the Galaxy mass is negligible, and a second one where the total LMC mass is introduced as a free parameter in the Hubble flow model. Using the SCM to fit the dynamics of nearby ($0.8\lesssim R/\mpc\lesssim 3$) galaxies under the assumption $M_{\rm LMC}\approx 0$ yields $M_{\rm MW,lhf}=0.8^{+0.4}_{-0.3}\times 10^{12}M_\odot$ (P14). If we adopt the results of Fig.~\ref{fig:mass_appr} at face value and assume $\Gamma= \Gamma_{\rm pseudo}\simeq 0.6$ we find a Milky Way virial mass that may be as low as $M_{\rm MW,vir}\sim M_{\rm MW,lhf}/1.25\sim 0.64\times 10^{12}M_\odot$. 

In contrast, P16 find that the LMC may be more massive than previously thought (see also Besla et al. 2012 and Belokurov \& Koposov 2016). Implementing the proper motions of the LMC measured recently from HST data in Equation~(\ref{eq:angmom2}) and fitting $M_{\rm LMC}$ and $M_{\rm MW, lfh}$ simultaneously to the same data set as in P14 returns a similar effective Milky Way mass, $M_{\rm MW,lhf}=1.04^{+0.42}_{-0.38}\times 10^{12}M_\odot$, while strongly favouring a massive LMC, $M_{\rm LMC}=0.25^{+0.09}_{-0.08}\times 10^{12}M_\odot$. These values yield $\Gamma_{\rm acc}\simeq 0.53$, and a total value of $\Gamma\sim 1.17$. Hence, combination of P16 constraints with Fig.~\ref{fig:mass_appr} suggests a virial mass $M_{\rm MW,vir}\sim M_{\rm MW,lhf}/1.20\sim 0.87\times 10^{12}M_\odot$, which is in excellent agreement with some of the values discussed in the introduction (e.g. Battaglia et al. 2005; Smith et al. 2007; Xue et al. 2008), but somewhat in tension with the high-mass bounds derived from the motion of Leo I (Sakamoto et al. 2003; Boylan-Kolchin et al. 2013).

The above estimates emphasize the practical challenges that must be faced when trying to infer the virial mass of a galaxy from the effective mass measured from the local Hubble flow. In general, our analysis indicates that extending the dynamical range of observational data sets to objects located beyond the virial radius of the Milky Way does {\it not} lead to more stringent constraints on $M_{\rm MW,vir}$. In fact, estimates of $M_{\rm MW,vir}$ based on the measured value of $M_{\rm MW,lhf}$ require additional information on the mass growth of our Galaxy, which is typically difficult to quantity. Unfortunately, the concept of virial mass based on an evolving overdensity threshold leads to pseudo-evolution, which adds further uncertainty to the relation between the two mass estimates.

\section{Summary}\label{sec:sum}
This paper uses twelve cosmological $N$-body simulations of Local Group systems (the {\sc Apostle} models, Sawala et al. 2016; Fattahi et al. 2016) to study the relation between the masses of the main galaxies enclosed within an evolving overdensity threshold ($M_{\rm vir,1}$ and $M_{\rm vir,2}$), the effective mass perturbing the Hubble flow ($M_{\rm lhf}$), and the timing-argument mass derived from the relative motion of the main galaxies ($M_{\rm tim}$). Our findings can be summarized as follows

\begin{itemize}

\item Despite the highly idealized assumptions on which it rests, the Spherical Collapse Model (SCM) provides an accurate description of the distance-velocity relation of galaxies within $d\lesssim R\lesssim 3\mpc$ from the main haloes, where $d$ is the current separation between the main galaxies.

\item The SCM establishes an exact equivalence between the three mass estimates, such that $M_{\rm lhf}=M_{\rm tim}=M_{\rm vir,1}+M_{\rm vir,2}$.
However, comparison against the {\sc Apostle} $N$-body simulations indicate that one of key SCM assumptions, namely that galaxies can be modelled as homogeneous over-densities with a finite size, $R_{\rm vir}$, leads to virial masses which systematically underestimate the mass derived from the local Hubble flow. We also find that the relation between the timing-argument mass and the combined virial mass of the pairs is strongly scattered, in agreement with previous studies.

\item On the other hand, we show that the local Hubble flow provides an unbiased constraint on the virial mass ratio, $f_m=M_{\rm lhf,1}/M_{\rm lhf,2}\approx M_{\rm vir,1}/M_{\rm vir,2}$, for all {\sc Apostle} haloes. Hence, combination of $f_m$ and $M_{\rm lhf}$ yields an estimate of the individual main galaxy masses without heuristic assumptions on the matter distribution or the equilibrium state of these systems.

\item Fitting Diemer \& Kravtsov (2014) density profile to the main {\sc Apostle} haloes shows that the individual galaxy masses inferred from the Hubble flow ($M_{{\rm lhf},i}$, for $i=1,2$) roughly correspond to the asymptotically-convergent (total) masses of the main haloes, and that the amount ``extra-virial'' mass ($\Delta M_i=M_{{\rm lhf},i}-M_{\rm vir,i}$) increases in galaxies with a slow growth rate. 

\item The uncertainties of the galaxy mass measurements are mainly driven by (i) the (unknown) tangential motion of the tracer haloes with respect to the Local Group barycentre, which increase the scatter in the observed Hubble flow, and (ii) the number of tracers in the fit.

\item In general, we find that estimates of $M_{\rm vir}$ based on the mass inferred from the dynamics of tracers at $R\gg R_{\rm vir}$ requires a priori information on the internal matter distribution {\it and} the growth rate of the galaxy, both of which are difficult to quantify.

\end{itemize}

\section{Acknowledgements}
We would like to thank the referee, Facundo G\'omez, for a careful and critical report. We are also indebted to Andrey Kravtsov for very useful discussions and comments on the draft. Our thanks also to members of the Apostle collaboration, Joop Schaye, Till Sawala, Carlos Frenk, and Julio Navarro. The Apostle simulation data used in this paper were created on the DiRAC Data Centric system at Durham University, operated
by the Institute for Computational Cosmology on behalf of the STFC DiRAC
HPC Facility (www.dirac.ac.uk), and also on resources provided by
WestGrid (www.westgrid.ca) and Compute Canada (www.computecanada.ca).
The DiRAC system was funded by BIS National E-infrastructure capital
grant ST/K00042X/1, STFC capital grants ST/H008519/1 and ST/K00087X/1,
STFC DiRAC Operations grant ST/K003267/1 and Durham University. DiRAC is
part of the National E-Infrastructure. This research has made use of
NASA's Astrophysics Data System.

{}

\newpage

\begin{table*}
\begin{tabular}{L{1.2cm}  | L{1.2cm} |  c c c c  c c  c c }
\hline
\hline
{\bf {\sc Apostle} Model}  \vspace{1mm} & {\bf Gal. ID} \vspace{1mm} & $M_{\rm vir}$ & $r_{\rm vir}$ & $c_{\rm vir}$ & $\Gamma$ & $M(<d)$ & $d$ & $M(<R_{\rm ta})$ & $R_{\rm ta}$\\
& & ($10^{12} M_\odot$) & ($\mpc$)& & &  ($10^{12} M_\odot$)  & ($\mpc$)   &($10^{12}M_\odot$) &($\mpc$) \\
\hline
\hline
\rowcolor{gray!40}AP-01 &393142 & 1.920 & 0.326 & 12.98 & 0.526 & 3.956 & 0.844 & 4.260 & 1.148\\[2ex]
\rowcolor{gray!40}     &420308 & 1.265 & 0.283 & 9.11 & 1.832 & 3.956 & 0.844 & 4.260 & 1.148\\[2ex]
                  AP-02 &433624 & 0.996 & 0.262 & 13.05& 0.791 & 3.044 & 0.820 & 3.563 & 1.213\\[2ex]
                       &438030 & 0.999 & 0.262 & 10.44 & 0.578 & 3.044 & 0.820 & 3.563 & 1.213\\[2ex]
\rowcolor{gray!40}AP-03 &388060 & 2.007 & 0.331 & 9.41 & 1.803 & 5.417 & 0.936 & 6.045 & 1.329\\[2ex]
\rowcolor{gray!40}     &392405 & 1.870 & 0.323 & 7.49 & 2.746 & 5.417 & 0.936 & 6.045 & 1.329\\[2ex]
                  AP-04 &396364 & 1.575 & 0.305 & 11.69 & 0.473 & 4.194 & 0.772 & 4.997 & 1.329\\[2ex]
                       &398846 & 1.706 & 0.313 & 8.91 & 2.585 & 4.194 & 0.772 & 4.997 & 1.329\\[2ex]
\rowcolor{gray!40}AP-05 &429125 & 1.103 & 0.271 & 10.39 & 0.823 & 3.366 & 0.812 & 4.612 & 1.116\\[2ex]
\rowcolor{gray!40}     &429774 & 1.046 & 0.266 & 12.05 & 0.152 & 3.366 & 0.812 & 4.612 & 1.116\\[2ex]
                  AP-06 &369074 & 2.777 & 0.368 & 10.47 & 0.567 & 5.267 & 0.915 & 6.455 & 1.447\\[2ex]
                       &412050 & 1.395 & 0.293 & 10.05 & 1.509 & 5.267 & 0.915 & 6.455 & 1.447\\[2ex]
\rowcolor{gray!40}AP-07 &330892 & 2.218 & 0.342 & 9.46 & 0.879 & 5.169 & 0.678 & 6.936 & 1.483\\[2ex]
\rowcolor{gray!40}     &330893 & 1.282 & 0.284 & 7.66 & 0.108 & 5.169 & 0.678 & 6.936 & 1.483\\[2ex]
                  AP-08 &356047 & 2.010 & 0.330 & 8.04 & 1.978 & 4.357 & 0.835 & 5.284 & 1.290\\[2ex]
                       &356048 & 0.749 & 0.238 & 8.78 & 0.558 & 4.357 & 0.835 & 5.284 & 1.290\\[2ex]
\rowcolor{gray!40}AP-09 &425486 & 1.152 & 0.274 &15.19 & 0.529 & 2.502 & 0.815 & 2.801 & 1.173\\[2ex]
\rowcolor{gray!40}     &456070 & 0.769 & 0.240 & 10.86 & 0.237 & 2.502 & 0.815 & 2.801 & 1.173\\[2ex]
                  AP-10 &368423 & 1.681 & 0.311 & 9.70 &-0.183 & 3.561 & 0.676 & 4.888 & 1.224\\[2ex]
                       &368424 & 1.140 & 0.274 & 11.85 & 1.949 & 3.561 & 0.676 & 4.888 & 1.224\\[2ex]
\rowcolor{gray!40}AP-11 &378313 & 1.173 & 0.276 & 12.52 & 0.928 & 3.895 & 0.780 & 4.504 & 1.395\\[2ex]
\rowcolor{gray!40}     &378314 & 0.970 & 0.259 & 9.57 & 1.036 & 3.895 & 0.780 & 4.504 & 1.395\\[2ex]
                  AP-12 &385279 & 1.179 & 0.277 &13.14 & 0.439 & 2.570 & 0.670 & 3.138 & 1.183\\[2ex]
                       &385280 & 0.784 & 0.241 & 10.02 & 1.892 & 2.570 & 0.670 & 3.138 & 1.183\\[2ex]
\hline
\hline
\end{tabular}
\caption{Quantities measured from the {\sc Apostle} $N$-body models. Note that the negative $\Gamma$-value of the halo 368423 results from massive substructures crossing repeatedly the virial radius of this sytem.}
\label{tab:nbody} 
\end{table*}

\begin{table*}
\begin{tabular}{L{1.2cm} | L{1.2cm} |  c c c c c | c}
\hline
\hline
& & & & {\bf Local Hubble flow}  & & & {\bf Timing argument}\\
\hline
{\bf {\sc Apostle} Model} & {\bf Gal. ID} \vspace{1mm} & $M_{\rm lhf}$ & $f_m$ & $\sigma_m$ & $M_1$ & $M_2$ & $M_{\rm tim}$\\
& &  ($10^{12} M_\odot$) & & (km/s) & ($10^{12}M_\odot$)   &($10^{12}M_\odot$) &($10^{12}M_\odot$) \\
\hline
\hline
\rowcolor{gray!40}AP-01 &393142 & $  3.22\pm  0.29$ & $  1.22\pm  0.19$ & $  28.4\pm   3.0$ & $  1.76\pm  0.20$ & $  1.46\pm  0.19$ &  2.84\\[2ex]
\rowcolor{gray!40}     &420308 & $  2.78\pm  0.29$ & $  1.08\pm  0.20$ & $  32.3\pm   3.3$ & $  1.43\pm  0.17$ & $  1.36\pm  0.21$ &  2.84\\[2ex]
                  AP-02 &433624 & $  3.23\pm  0.70$ & $  1.04\pm  0.67$ & $  66.3\pm   6.9$ & $  1.51\pm  0.46$ & $  1.72\pm  0.63$ &  3.64\\[2ex]
                       &438030 & $  3.86\pm  0.68$ & $  3.77\pm  2.02$ & $  59.3\pm   6.0$ & $  2.90\pm  0.66$ & $  0.93\pm  0.37$ &  3.64\\[2ex]
\rowcolor{gray!40}AP-03 &388060 & $  4.83\pm  0.61$ & $  1.05\pm  0.26$ & $  50.3\pm   5.0$ & $  2.44\pm  0.41$ & $  2.39\pm  0.42$ &  3.65\\[2ex]
\rowcolor{gray!40}     &392405 & $  4.49\pm  0.51$ & $  1.42\pm  0.33$ & $  45.7\pm   4.6$ & $  2.60\pm  0.37$ & $  1.88\pm  0.33$ &  3.65\\[2ex]
                  AP-04 &396364 & $  4.99\pm  0.31$ & $  1.55\pm  0.21$ & $  24.5\pm   2.5$ & $  3.02\pm  0.22$ & $  1.97\pm  0.22$ &  2.41\\[2ex]
                       &398846 & $  4.35\pm  0.22$ & $  1.00\pm  0.11$ & $  19.6\pm   2.0$ & $  2.17\pm  0.17$ & $  2.18\pm  0.15$ &  2.41\\[2ex]
\rowcolor{gray!40}AP-05 &429125 & $  2.77\pm  0.58$ & $  1.55\pm  0.76$ & $  64.5\pm   6.6$ & $  1.60\pm  0.39$ & $  1.17\pm  0.40$ &  3.77\\[2ex]
\rowcolor{gray!40}     &429774 & $  2.74\pm  0.61$ & $  1.07\pm  0.53$ & $  65.7\pm   6.6$ & $  1.38\pm  0.48$ & $  1.37\pm  0.34$ &  3.77\\[2ex]
                  AP-06 &369074 & $  5.74\pm  0.34$ & $  2.12\pm  0.29$ & $  26.6\pm   2.7$ & $  3.89\pm  0.28$ & $  1.86\pm  0.20$ &  3.27\\[2ex]
                       &412050 & $  6.29\pm  0.38$ & $  0.67\pm  0.09$ & $  28.6\pm   2.9$ & $  2.50\pm  0.25$ & $  3.79\pm  0.31$ &  3.27\\[2ex]
\rowcolor{gray!40}AP-07 &330892 & $  6.59\pm  0.65$ & $  2.26\pm  0.72$ & $  44.7\pm   4.5$ & $  4.48\pm  0.50$ & $  2.12\pm  0.49$ &  5.77\\[2ex]
\rowcolor{gray!40}     &330893 & $  6.36\pm  0.60$ & $  0.44\pm  0.10$ & $  40.8\pm   4.2$ & $  1.94\pm  0.41$ & $  4.41\pm  0.41$ &  5.77\\[2ex]
                  AP-08 &356047 & $  4.39\pm  0.45$ & $  2.14\pm  0.47$ & $  41.4\pm   4.3$ & $  2.96\pm  0.38$ & $  1.43\pm  0.24$ &  5.84\\[2ex]
                       &356048 & $  4.12\pm  0.48$ & $  0.44\pm  0.11$ & $  42.1\pm   4.4$ & $  1.25\pm  0.30$ & $  2.87\pm  0.35$ &  5.84\\[2ex]
\rowcolor{gray!40}AP-09 &425486 & $  3.12\pm  0.18$ & $  1.79\pm  0.25$ & $  18.0\pm   1.9$ & $  1.99\pm  0.14$ & $  1.13\pm  0.13$ &  1.95\\[2ex]
\rowcolor{gray!40}     &456070 & $  3.28\pm  0.19$ & $  0.77\pm  0.10$ & $  19.4\pm   2.0$ & $  1.42\pm  0.14$ & $  1.86\pm  0.15$ &  1.95\\[2ex]
                  AP-10 &368423 & $  3.64\pm  0.45$ & $  3.83\pm  1.36$ & $  40.4\pm   4.1$ & $  2.83\pm  0.32$ & $  0.81\pm  0.25$ &  2.46\\[2ex]
                       &368424 & $  3.63\pm  0.38$ & $  0.45\pm  0.10$ & $  36.3\pm   3.7$ & $  1.13\pm  0.23$ & $  2.50\pm  0.27$ &  2.46\\[2ex]
\rowcolor{gray!40}AP-11 &378313 & $  5.16\pm  0.56$ & $  1.37\pm  0.31$ & $  44.5\pm   4.6$ & $  2.94\pm  0.38$ & $  2.22\pm  0.41$ &  4.92\\[2ex]
\rowcolor{gray!40}     &378314 & $  5.61\pm  0.55$ & $  0.89\pm  0.18$ & $  39.8\pm   4.0$ & $  2.62\pm  0.43$ & $  3.00\pm  0.35$ &  4.92\\[2ex]
                  AP-12 &385279 & $  3.22\pm  0.24$ & $  0.68\pm  0.09$ & $  23.5\pm   2.4$ & $  1.92\pm  0.17$ & $  1.29\pm  0.15$ &  1.64\\[2ex]
                       &385280 & $  3.35\pm  0.26$ & $  2.00\pm  0.35$ & $  25.5\pm   2.8$ & $  1.14\pm  0.17$ & $  2.22\pm  0.19$ &  1.64\\[2ex]
\hline
\hline
\end{tabular}
\caption{Quantities inferred from the dynamics of haloes between 1--3$\mpc$ from the barycentre of the Local Group analogues ({\it local Hubble flow}), and from the relative motion betwen the main galaxies of the groups ({\it classical timing argument}).}
\label{tab:fits} 
\end{table*}

\end{document}